\documentclass[prb,twocolumn,showpacs]{revtex4}
\usepackage{graphicx} 
\usepackage{amsmath,amsthm,amssymb,mathrsfs}
\usepackage{bm}

\begin{document}

\title{Spin-current driven spontaneous coupling of ferromagnets}

\author{Tomohiro Taniguchi}
 \affiliation{
 National Institute of Advanced Industrial Science and Technology (AIST), Spintronics Research Center, Tsukuba, Ibaraki 305-8568, Japan 
 }

 \begin{abstract}
{ 
A theoretical framework is proposed for the spin-current driven synchronized self-oscillations in ferromagnets in the spin Hall geometry. 
The spin current generated by the spin Hall effect in a bottom nonmagnetic heavy metal excites 
a self-oscillation of the magnetization in an attached ferromagnet through spin-transfer effect. 
The spin current simultaneously creates spin accumulation inside the ferromagnet. 
Therefore, when the top surfaces of two ferromagnets are connected by a nonmagnetic material having a long spin diffusion length, 
another spin current flows according to the gradient of the spin accumulations between the ferromagnets. 
This additional spin current excites an additional spin torque leading to a coupled motion of the magnetizations. 
This coupling mechanism comes purely from spin degree of freedom in the system without using electric and/or magnetic interactions. 
The additional spin torque acts as a repulsive force between the magnetizations, and prefers an antiphase synchronization between the oscillators. 
The phase difference in a synchronized state is determined 
by the competition between this additional spin torque and spin pumping. 
Eventually, either an in-phase or antiphase synchronization is spontaneously excited in the individual ferromagnets, 
depending on the current magnitude. 
These conclusions are obtained by deriving the theoretical formula of the additional spin torque from the diffusive spin transport theory 
and solving the equation of motion of the magnetizations both numerically and analytically. 
}
 \end{abstract}

 \maketitle

% ================================================================================================================================================================================= %

% ===================================================================================================================================================================================== %

\section{Introduction}
\label{sec:Introduction}

Current driven magnetization dynamics in ferromagnets has been attracting much attention in the field of spintronics [\onlinecite{slonczewski96,berger96,kiselev03,rippard04,houssameddine07}]. 
In particular, an excitation of the coupled or forced dynamics of the magnetizations, 
such as synchronization of self-oscillations in spin torque oscillators (STOs), is currently an exciting topic. 
This is because it has a possibility in enhancing emission power of practical devices, such as microwave generator and magnetic sensors, 
and applicability to new devices such as phased array and brain-inspired computing [\onlinecite{locatelli14,grollier16,kudo17,torrejon17}]. 
The coupled dynamics is excited as a result of electric and/or magnetic interaction between the ferromagnets. 
Several mechanisms of coupling have been proposed theoretically and/or demonstrated experimentally, 
such as spin wave propagation [\onlinecite{kaka05,mancoff05,houshang16,urazhdin16,awad17}], electric current injection and/or feedback [\onlinecite{rippard05,kudo06,zhou09,khalsa15,tsunegi16,bhuktare17}], 
microwave field [\onlinecite{urazhdin10}], stochastic noise in current [\onlinecite{nakada12}], and dipole interaction [\onlinecite{locatelli15,chen16PRB}]. 
Each coupling mechanism has interesting characteristics. 
As an example, the dipole interaction can excite spontaneous synchronization without adding interconnections between STOs, 
while the number of STOs to be synchronized is restricted due to the spatial decay of the interaction. 
On the other hand, the electric current injection can lead to a long-range coupling owing to the conservation law of the current, but makes the circuit complicated. 
Moreover, every mechanism has the possibility leading to different types of synchronizations, depending on the experimental setup. 
For example, the electric coupling results in either in-phase or antiphase synchronization, depending on the method connecting the oscillators [\onlinecite{taniguchi18}]. 
Therefore, the investigation of a new coupling mechanism between STOs and clarification of its role on the synchronization provide rich physical insights for both magnetism and nonlinear science. 

% ===================================================================================================================================================================================== %

An excitation of spin-current driven coupled motion of magnetizations is a relatively new and interesting research target. 
%whereas the excitation of the self-oscillation in a single STO by pure spin current has been demonstrated in the spin Hall \cite{liu12} and nonlocal \cite{urazhdin16} geometries. 
The coupled magnetization dynamics through spin current has been investigated mainly in ferromagnetic resonances (FMR) [\onlinecite{tserkovnyak03,tserkovnyak03JAP,heinrich03,takahashi14,skarsvag14,chiba15}]. 
However, a harmonic oscillation excited by an oscillating force, as in the case of FMR, should be distinguished by 
a self-oscillation excited by a direct force [\onlinecite{pikovsky03}]. 
A synchronization of the self-oscillation between STOs through spin current has not been fully investigated yet. 
The spin current decays within a characteristic length scale called spin diffusion length. 
Thus, the STOs should be connected with each other within a distance shorter than the spin diffusion length. 
It has been experimentally confirmed that several nonmagnetic metals, such as Cu and Al, have the spin diffusion length longer than one hundred nanometers [\onlinecite{bass07}]. 
Connecting STOs by such materials within a distance shorter than the spin diffusion length is expected. 
Therefore, it is of great interest to develop a model of coupling between ferromagnets through spin current and investigate synchronized magnetization dynamics. 

% ===================================================================================================================================================================================== %

In this work, we propose a theoretical model embodying spin-current driven synchronization between STOs in spin Hall geometry. 
The coupling mechanism in this work comes purely from spin degree of freedom in the system without using electric and/or magnetic interactions. 
We consider two STOs placed onto different nonmagnetic heavy metals. 
The spin currents generated by the spin Hall effects in the bottom electrodes excite the self-oscillations in the STOs through the spin-transfer effect. 
At this stage, two STOs oscillate independently. 
Here, the spin currents simultaneously create spin accumulation in the ferromagnets. 
Next, the top surfaces of the STOs are connected by another nonmagnetic metal having a long spin diffusion length. 
Then, additional spin currents flow through the top connector according to the gradient of the spin accumulation and due to the spin pumping mechanism. 
These spin currents excite additional spin torques resulting in a coupled motion of magnetizations. 
As a result, a phase synchronization is then spontaneously excited between the STOs. 
The phase difference between the STOs in the synchronized state is determined according to the competition between two coupling mechanisms; 
one originates from the spin current generated by the gradient of the spin accumulation whereas the other comes from the spin pumping effect. 
We find these results by developing the theoretical model of the additional spin torque from the diffusive spin transport theory 
and solving the equation of motion of the magnetizations both numerically and analytically. 

% ===================================================================================================================================================================================== %

This paper is organized as follows. 
In Sec. \ref{sec:Spin torque formula from spin transport theory}, we present a description of the system in this study. 
We first derive the distribution of the spin accumulation in a single STO in the presence of the spin Hall effect. 
Secondly, we consider connecting the top surfaces of two independent STOs, 
and investigate the coupling due to spin current flowing caused by the gradient of the spin accumulation and spin pumping in the connector. 
%The theoretical formulas of spin torques arising from the injections from both the bottom and top nonmagnets are then derived. 
In Sec. \ref{sec:Coupled magnetization dynamics}, we study the coupled motion of the magnetizations by solving the Landau-Lifshitz-Gilbert (LLG) equation. 
In Sec. \ref{sec:Role of coupling spin torque}, the role of the spin current generated by the gradient of the spin accumulation on the coupled motion of the magnetization 
is studied both numerically and analytically. 
Section \ref{sec:Summary} shows the conclusions of this work. 

% ===================================================================================================================================================================================== %

% ===================================================================================================================================================================================== %

% ===================================================================================================================================================================================== %

% ===================================================================================================================================================================================== %

\begin{figure}%[p]
\centerline{\includegraphics[width=1.0\columnwidth]{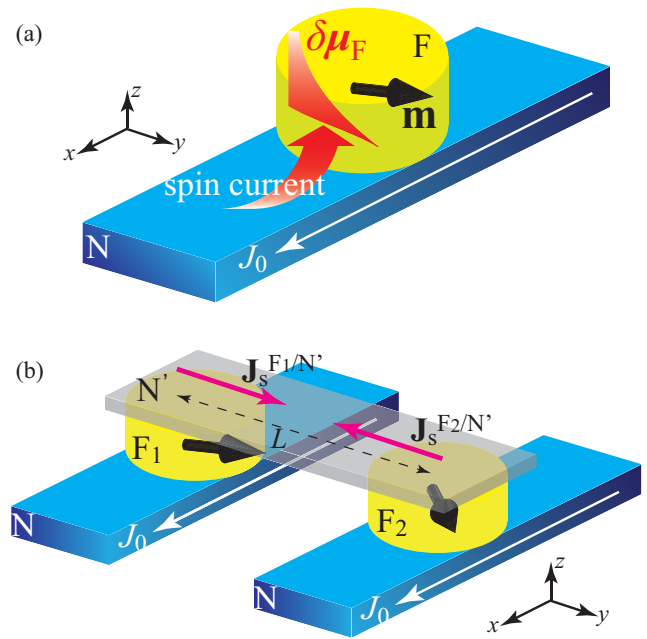}}%\vspace{-3.0ex}
\caption{
         (a) Schematic view of a single STO. 
             A ferromagnet F is placed onto a nonmagnet N. 
             The electric current density $J_{0}$ flowing in the nonmagnet is converted to a pure spin current injected into the ferromagnet by the spin Hall effect. 
             The spin current creates the spin accumulation $\delta\bm{\mu}_{\rm F}$ in the ferromagnet. 
         (b) Schematic view of the system having two STOs. 
             Another nonmagnet N${}^{\prime}$ is connected to the top surface of the ferromagnets, F${}_{\ell}$ ($\ell=1,2$). 
             Spin currents driven by the spin accumulations and spin pumping flow in the N${}^{\prime}$ layer. 
         \vspace{-3ex}}
\label{fig:fig1}
\end{figure}

% ===================================================================================================================================================================================== %

% ===================================================================================================================================================================================== %

% ===================================================================================================================================================================================== %

\section{Spin torque formula from spin transport theory}
\label{sec:Spin torque formula from spin transport theory}

In this section, we derive the spin torque formula in the spin Hall geometry from the diffusive spin transport theory. 

% ===================================================================================================================================================================================== %

\subsection{System description}
\label{sec:System description}

Let us first consider the injection of spin current into a single ferromagnet by the spin Hall effect. 
The system we consider is schematically shown in Fig. \ref{fig:fig1}(a). 
The electric current $J_{0}=\sigma_{\rm N}E_{x}$ flowing in the nonmagnet N along the $x$ direction is converted to a spin current flowing into the $z$ direction by the spin Hall effect, 
where $\sigma_{\rm N}$ is the conductivity of the nonmagnet N and $E_{x}$ is the electric field. 
The spin current creates the spin accumulation in the metallic ferromagnet F. 
Here, we define the spin accumulation $\delta\bm{\mu}_{\rm F}$ in the ferromagnet as $\delta\bm{\mu}_{\rm F}=[(\bar{\mu}_{{\rm F},\uparrow}-\bar{\mu}_{{\rm F},\downarrow})/2] \mathbf{m}$, 
where $\bar{\mu}_{{\rm F},s}$ ($s=\uparrow,\downarrow$) is the electrochemical potential of spin-$s$ electrons, 
and $\mathbf{m}$ is the unit vector pointing in the magnetization direction of the ferromagnet. 
%Here, we assume that the penetration depth of transverse spin current in the ferromagnet is sufficiently short, 
%and therefore, the spin accumulation in the ferromagnet is parallel to the magnetization \cite{slonczewski96,brataas00,brataas01,stiles02,zhang02,zhang04,zwierzycki05,taniguchi08,ghosh12}. 
The spin accumulation in the ferromagnet obeys the diffusion equation [\onlinecite{valet93,takahashi08}] 
\begin{equation}
  \frac{\partial^{2}}{\partial z^{2}}
  \delta
  \bm{\mu}_{{\rm F}}
  =
  \frac{\delta\bm{\mu}_{\rm F}}{\lambda_{\rm F}^{2}},
  \label{eq:diffusion_equation}
\end{equation}
where $\lambda_{\rm F}$ is the spin diffusion length in the ferromagnet. 

The solution of the spin accumulation is determined by identifying the spin currents at the boundaries. 
Let us denote that spin current density at the F/N interface ($z=0$) flowing in the positive $z$ direction, i.e., from the N to F layer, as $\mathbf{J}_{\rm s}^{\rm F/N}$. 
In both the ferromagnet and the nonmagnet, the spin currents are zero at the outer boundaries, $z=-d_{\rm N}$ and $z=d_{\rm F}$, 
where $d_{\rm N}$ and $d_{\rm F}$ are the thicknesses of the bottom nonmagnet and ferromagnet, respectively. 
Note that the spin currents in the ferromagnet and nonmagnet are given by 
\begin{equation}
  J_{{\rm s}i,{\rm F}}
  =
  -\frac{\hbar \sigma_{\rm F}}{2e^{2}}
  \partial_{i}
  \delta
  \mu_{\rm F}
  -
  \frac{\hbar \beta \sigma_{\rm F}}{2e^{2}}
  \partial_{i}
  \bar{\mu}_{\rm F},
  \label{eq:spin_current_F}
\end{equation}
\begin{equation}
  J_{{\rm s}i\alpha,{\rm N}}
  =
  -\frac{\hbar \sigma_{\rm N}}{2e^{2}}
  \partial_{i}
  \delta
  \mu_{{\rm N},\alpha}
  -
  \frac{\hbar \vartheta \sigma_{\rm N}}{2e^{2}}
  \epsilon_{i\alpha j}
  \partial_{j}
  \bar{\mu}_{\rm N},
  \label{eq:spin_current_N}
\end{equation}
where $\sigma_{\rm F}$ is the conductivity of the ferromagnet, and $\beta$ is its spin polarization. 
The spin Hall angle in the bottom nonmagnet is $\vartheta$. 
The suffixes $i$ and $j$ represent the spatial direction in the real space, 
whereas $\alpha$ represents the direction of the spin polarization. 
Note that the suffix $\alpha$ is not added to the spin current in the ferromagnet 
because the spin polarization in the ferromagnet is assumed to be parallel to the magnetization. 
The Levi-Civita asymmetric tensor is $\epsilon_{ijk}$ with $\epsilon_{123}=+1$. 
The electrochemical potential is denoted as $\bar{\mu}=(\bar{\mu}_{\uparrow}+\bar{\mu}_{\downarrow})/2$. 
Then, the solutions of the spin accumulation in the ferromagnet and nonmagnet are given by 
\begin{equation}
  \delta
  \bm{\mu}_{\rm F}
  =
  \frac{2e^{2} \lambda_{\rm F}}{\hbar (1-\beta^{2}) \sigma_{\rm F} \sinh(d_{\rm F}/\lambda_{\rm F})}
  \mathbf{m}
  \cdot
  \mathbf{J}_{\rm s}^{\rm F/N}
  \cosh
  \left(
    \frac{z-d_{\rm F}}{\lambda_{\rm F}}
  \right)
  \mathbf{m},
  \label{eq:spin_accumulation_F}
\end{equation}
\begin{equation}
\begin{split}
  \delta
  \bm{\mu}_{\rm N}
  =&
  \frac{2e^{2} \lambda_{\rm N}}{\hbar \sigma_{\rm N} \sinh(d_{\rm N}/\lambda_{\rm N})}
  \left[
    -\frac{\hbar \vartheta \sigma_{\rm N}}{2e}
    E_{x}
    \cosh
    \left(
      \frac{z}{\lambda_{\rm N}}
    \right)
    \mathbf{e}_{y}
  \right.
\\
  &
  \left.
    -
    \left(
      \mathbf{J}_{\rm s}^{\rm F/N}
      -
      \frac{\hbar \vartheta \sigma_{\rm N}}{2e}
      E_{x}
      \mathbf{e}_{y}
    \right)
    \cosh
    \left(
      \frac{z+d_{\rm N}}{\lambda_{\rm N}}
    \right)
  \right].
  \label{eq:spin_accumulation_N}
\end{split}
\end{equation}

The spin current at the F/N boundary is given by the circuit theory [\onlinecite{brataas01}] as 
\begin{equation}
\begin{split}
  \mathbf{J}_{\rm s}^{\rm F/N}
  =&
  -\frac{1}{2\pi S}
  \left[
    \frac{(1-p_{g}^{2}) g}{2}
    \mathbf{m}
    \cdot
    \left(
      \delta
      \bm{\mu}_{\rm F}
      -
      \delta
      \bm{\mu}_{\rm N}
    \right)
    \mathbf{m}
  \right.
\\
  &
  \left.
    -
    g_{\rm r}
    \mathbf{m}
    \times
    \left(
      \delta
      \bm{\mu}_{\rm N}
      \times
      \mathbf{m}
    \right)
    -
    g_{\rm i}
    \delta
    \bm{\mu}_{\rm N}
    \times
    \mathbf{m}
  \right],
  \label{eq:spin_current_FN}
\end{split}
\end{equation}
where $S$ is the cross section area of the F/N boundary. 
The interface conductance $g$ with its spin polarization $p_{g}$ is related to the interface resistance $r$ via $r=(h/e^{2})S/g$. 
The real and imaginary parts of the mixing conductance are denoted as $g_{\rm r}$ and $g_{\rm i}$, respectively. 
Substituting Eqs. (\ref{eq:spin_accumulation_F}) and (\ref{eq:spin_accumulation_N}) into Eq. (\ref{eq:spin_current_FN}), 
we find that Eq. (\ref{eq:spin_current_FN}) can be rewritten as [\onlinecite{chen13,taniguchi16PRB}]
\begin{equation}
\begin{split}
  \mathbf{J}_{\rm s}^{\rm F/N}
  =&
  \frac{\hbar \vartheta \sigma_{\rm N} g^{*}}{2e g_{\rm N}}
  m_{y}
  E_{x}
  \tanh
  \left(
    \frac{d_{\rm N}}{2 \lambda_{\rm N}}
  \right)
  \mathbf{m}
\\
  &+
  \frac{\hbar \sigma_{\rm N}}{2e}
  E_{x}
  \left[
    \vartheta_{\rm R}
    \mathbf{m}
    \times
    \left(
      \mathbf{e}_{y}
      \times
      \mathbf{m}
    \right)
    +
    \vartheta_{\rm I}
    \mathbf{e}_{y}
    \times
    \mathbf{m}
  \right]
  \label{eq:spin_current_FN_sol}
\end{split}
\end{equation}
where we introduce the following notations, 
\begin{equation}
  \frac{1}{g^{*}}
  =
  \frac{2}{(1-p_{g}^{2})g}
  +
  \frac{1}{g_{\rm F} \tanh(d_{\rm F}/\lambda_{\rm F})}
  +
  \frac{1}{g_{\rm N} \tanh(d_{\rm N}/\lambda_{\rm N})},
\end{equation}
\begin{equation}
  \frac{g_{\rm F}}{S}
  =
  \frac{h (1-\beta^{2}) \sigma_{\rm F}}{2e^{2}\lambda_{\rm F}},
\end{equation}
\begin{equation}
  \frac{g_{\rm N}}{S}
  =
  \frac{h \sigma_{\rm N}}{2e^{2} \lambda_{\rm N}}.
\end{equation}
\begin{equation}
  \vartheta_{\rm R(I)}
  =
  \vartheta
  \tanh
  \left(
    \frac{d_{\rm N}}{2\lambda_{\rm N}}
  \right)
  {\rm Re}
  \left(
    {\rm Im}
  \right)
  \frac{g_{\rm r}+ i g_{\rm i}}{g_{\rm N} + (g_{\rm r} + i g_{\rm i}) \coth(d_{\rm N}/\lambda_{\rm N})}.
\end{equation}
For typical ferromagnetic/nonmagnetic metallic interface, $g_{\rm r} \gg |g_{\rm i}|$ [\onlinecite{zwierzycki05}]. 
Therefore, in the following, we neglect the terms related to $g_{\rm i}$. 

The absorption of the transverse spin current at the F/N interface leads to an excitation of spin torque acting on the magnetization in the ferromagnet. 
We denote the spin current at the F/N interface flowing from the ferromagnet to the nonmagnet as $\mathbf{J}_{\rm s}^{\rm F \to N}=-\mathbf{J}_{\rm s}^{\rm F/N}$. 
By using Eq. (\ref{eq:spin_current_FN_sol}), 
the conventional spin Hall torque generated from the spin current injected from the bottom nonmagnet is given by 
\begin{equation}
\begin{split}
  \mathbf{T}^{(1)}
  &=
  \frac{\gamma}{Md_{\rm F}}
  \mathbf{m}
  \times 
  \left(
    \mathbf{J}_{\rm s}^{\rm F\to N}
    \times
    \mathbf{m}
  \right)
\\
  &=
  -\frac{\gamma \hbar \vartheta_{\rm R} J_{0}}{2eMd_{\rm F}}
  \mathbf{m}
  \times
  \left(
    \mathbf{e}_{y}
    \times
    \mathbf{m}
  \right). 
  \label{eq:STT_1}
\end{split}
\end{equation}
We note that the spin torque in Eq. (\ref{eq:STT_1}) is expressed in terms of the current density, $J_{0}$. 
%We assumes the same sections between the ferromagnet and nonmagnet, for simplicity.
Throughout this paper, we use the current density $J_{0}$, and not a current which is a product of $J_{0}$ and the cross-section area of the nonmagnet in the $yz$ plane. 
This is, from the viewpoint of simplicity, it is preferable to avoid to including geometrical factor [\onlinecite{fert01}] related to the different size between the ferromagnet and nonmagnet, 
although the experimental results are often expressed in terms of current. 

In the next section, we consider connecting the ferromagnets by adding another electrode on their top surfaces. 
In this case, it is necessary to evaluate the spin accumulation at the top surface of the ferromagnet 
to understand the role of the connection on the magnetization dynamics. 
Substituting Eq. (\ref{eq:spin_current_FN_sol}) into Eq. (\ref{eq:spin_accumulation_F}), 
the solution of the spin accumulation in the ferromagnet is given by 
\begin{equation}
\begin{split}
  \delta
  \bm{\mu}_{{\rm F}}
  =&
  e \vartheta^{*}
  \lambda_{\rm F}
  E_{x}
  m_{y}
  \cosh
  \left(
    \frac{z-d_{\rm F}}{\lambda_{\rm F}}
  \right) 
  \mathbf{m},
  \label{eq:spin_accumulation}
\end{split}
\end{equation}
where $\vartheta^{*}$ is defined as 
\begin{equation}
  \vartheta^{*}
  =
  \vartheta
  \frac{\sigma_{\rm N} g^{*} \tanh[d_{\rm N}/(2 \lambda_{\rm N})]}{(1-\beta^{2}) \sigma_{\rm F} g_{\rm N} \sinh(d_{\rm F}/\lambda_{\rm F})}.
\end{equation}
As schematically shown in Fig. \ref{fig:fig1}(a) and described by Eq. (\ref{eq:spin_accumulation}), 
the spin accumulation in the ferromagnet is maximized at the F/N interface, 
and decreases exponentially from the interface. 
We note that the spin accumulation at the outer boundary, $z=d_{\rm F}$, is finite. 
It should also be emphasized that the magnitude and polarized-direction of the spin accumulation depend on the magnetization direction 
through the term $m_{y}\mathbf{m}$ in Eq. (\ref{eq:spin_accumulation}). 

% ===================================================================================================================================================================================== %

\subsection{Spin current in nonmagnetic connector}
\label{sec:Spin current in nonmagnetic connector}

Now let us consider two STOs and assume placing another nonmagnet N${}^{\prime}$ onto the top surface of the ferromagnets, as shown in Fig. \ref{fig:fig1}(b). 
In the following, we add the suffix $\ell=1,2$ to the quantities related to the F${}_{\ell}$ layer to distinguish the ferromagnets. 
We assume for simplicity that the cross section area of the ferromagnet in the $xy$ plane and that of the nonmagnetic connector in the $xz$ plane are the same.

When two STOs are connected by the nonmagnet N${}^{\prime}$, 
spin currents are driven in the nonmagnet N${}^{\prime}$ according to the drop of the spin accumulation at the F${}_{\ell}$/N${}^{\prime}$ interface 
and the gradient of the spin accumulation inside the connector. 
The spin current at the interface is described by 
\begin{equation}
  \mathbf{J}_{\rm s}^{{\rm F}_{\ell} \to {\rm N^{\prime}}}
  =
  \frac{1}{2\pi S^{\prime}}
  \left[
    \frac{(1-p_{g}^{\prime\ 2})g^{\prime}}{2}
    \mathbf{m}_{\ell}
    \cdot
    \left(
      \delta
      \bm{\mu}_{{\rm F}_{\ell}}
      -
      \delta
      \bm{\mu}_{\rm N^{\prime}}
    \right)
    \mathbf{m}_{\ell}
    -
    g_{\rm r}^{\prime}
    \mathbf{m}_{\ell}
    \times
    \left(
      \delta
      \bm{\mu}_{\rm N^{\prime}}
      \times
      \mathbf{m}_{\ell}
    \right)
  \right], 
  \label{eq:spin_current_FN_appendix}
\end{equation}
where $\delta\bm{\mu}_{\rm N^{\prime}}$ is the spin accumulation in the connector N${}^{\prime}$. 
We note that Eq. (\ref{eq:spin_current_FN_appendix}) is basically identical to Eq. (\ref{eq:spin_current_FN}) 
except for the fact that Eq. (\ref{eq:spin_current_FN_appendix}) is defined at the F${}_{\ell}$/N${}^{\prime}$ interface 
whereas Eq. (\ref{eq:spin_current_FN}) is defined at the F${}_{\ell}$/N interface. 
The sign difference on the right hand sides of Eqs. (\ref{eq:spin_current_FN}) and (\ref{eq:spin_current_FN_appendix}) 
is due to the fact that the spin current defined by Eq. (\ref{eq:spin_current_FN}) flows from the nonmagnet N to the ferromagnet F${}_{\ell}$, 
whereas that defined by Eq. (\ref{eq:spin_current_FN_appendix}) flows from the ferromagnet F${}_{\ell}$ to the nonmagnetic connector N${}^{\prime}$. 
In Eq. (\ref{eq:spin_current_FN_appendix}), we add the prime symbols to the quantities related to the F${}_{\ell}$/N${}^{\prime}$ interface 
to distinguish them from those defined at the F${}_{\ell}$/N interface in Eq. (\ref{eq:spin_current_FN}). 
We assume that the quantities related to the interface resistance at 
F${}_{1}$/N${}^{\prime}$ are identical to those at F${}_{2}$/N${}^{\prime}$ interface. 

% ================================================================================================================================================================================= %

Using Eq. (\ref{eq:spin_accumulation}), Eq. (\ref{eq:spin_current_FN_appendix}) is rewritten as 
\begin{equation}
\begin{split}
  \mathbf{J}_{\rm s}^{{\rm F}_{\ell} \to {\rm N^{\prime}}}
  =&
  \frac{-1}{2\pi S^{\prime}}
  \left[
  \frac{(1-p_{\rm g}^{\prime\ 2})g^{\prime}}{2}
    \left(
      \mathbf{m}_{\ell}
      \cdot
      \delta
      \bm{\mu}_{\rm N^{\prime}}
    \right)
    \mathbf{m}_{\ell}
  \right.
\\
  &
  \left.
    +
    g_{\rm r}^{\prime}
    \mathbf{m}_{\ell}
    \times
    \left(
      \delta
      \bm{\mu}_{\rm N^{\prime}}
      \times
      \mathbf{m}_{\ell}
    \right)
  \right]
\\
  &+
  \frac{(1-p_{\rm g}^{\prime\ 2})g^{\prime}}{4\pi S^{\prime}}
  e \vartheta^{*}
  \lambda_{\rm F}
  E_{x}
  m_{\ell y}
  \mathbf{m}_{\ell}. 
  \label{eq:spin_current_FN_appendix_1}
\end{split}
\end{equation}
We note that the emission of the spin current given by Eq. (\ref{eq:spin_current_FN_appendix_1}) results in an additional spin torque acting on the magnetization $\mathbf{m}_{\ell}$. 
The last term on the right-hand side of Eq. (\ref{eq:spin_current_FN_appendix_1}), 
\begin{equation}
  \mathbf{J}_{\rm s}^{{\rm SHE}(\ell)}
  =
  \frac{(1-p_{\rm g}^{\prime\ 2})g^{\prime}}{4\pi S^{\prime}}
  e \vartheta^{*}
  \lambda_{\rm F}
  E_{x}
  m_{\ell y}
  \mathbf{m}_{\ell}, 
  \label{eq:spin_current_SHE}
\end{equation}
is a source term of this additional spin torque. 
There are several approaches to derive the spin torque formula from Eq. (\ref{eq:spin_current_FN_appendix_1}). 

% ================================================================================================================================================================================= %

One is to solve the diffusion equation of $\delta\bm{\mu}_{\rm N^{\prime}}$. 
The spin accumulation in the nonmagnetic connector obeys the diffusion equation with the spin diffusion length $\lambda_{\rm N^{\prime}}$, 
and is given by 
\begin{equation}
\begin{split}
  \delta
  \bm{\mu}_{\rm N^{\prime}}
  =
  \frac{2e^{2} \lambda_{\rm N^{\prime}}}{\hbar \sigma_{\rm N^{\prime}} \sinh(L/\lambda_{\rm N^{\prime}})}
  &
  \left[
    \mathbf{J}_{\rm s}^{{\rm F_{1}} \to {\rm N^{\prime}}}
    \cosh
    \left(
      \frac{y-L}{\lambda_{\rm N^{\prime}}}
    \right)
  \right.
\\
  &
  \left.
    +
    \mathbf{J}_{\rm s}^{{\rm F_{2}} \to {\rm N^{\prime}}}
    \cosh
    \left(
      \frac{y}{\lambda_{\rm N^{\prime}}}
    \right)
  \right],
  \label{eq:spin_accumulation_N_p_appendix}
\end{split}
\end{equation}
where $\sigma_{\rm N^{\prime}}$ is the conductivity of the nonmagnetic connector. 
We assume that F${}_{1}$ and F${}_{2}$ layers locate at $y=0$ and $y=L$, respectively. 
Substituting Eq. (\ref{eq:spin_accumulation_N_p_appendix}) into Eq. (\ref{eq:spin_current_FN_appendix_1}), 
the spin current at the F${}_{\ell}$/N${}^{\prime}$ interface, $\mathbf{J}_{\rm s}^{{\rm F}_{\ell} \to {\rm N^{\prime}}}$, 
will be expressed as a function of the source term. 
This approach has been used in, for example, Refs. [\onlinecite{chiba15,chen13}] for a situation, 
where a ferromagnet is an insulator, and therefore, $g^{\prime} \to 0$. 
The coupling spin torque derived by this approach will explicitly depend on the length of the nonmagnetic connector $L$, 
similar to Eq. (\ref{eq:STT_2}).

% ================================================================================================================================================================================= %

Another approach is to assume a ballistic transport in the nonmagnetic connector N${}^{\prime}$. 
In this case, the spin current is conserved in the nonmagnetic connector, i.e., $\mathbf{J}_{\rm s}^{{\rm F}_{1} \to {\rm N^{\prime}}}+\mathbf{J}_{\rm s}^{{\rm F}_{2} \to {\rm N^{\prime}}}=\bm{0}$. 
Then again, the spin current at the F${}_{\ell}$/N${}^{\prime}$ interface will be expressed as a function of the source term. 
The spin torque formula derived by this approach, however, does not include the length of the nonmagnetic connector 
due to the assumption of the conservation law of the spin current. 
The derivations of the spin torque formulas based on this approach were developed in, for example, Refs. [\onlinecite{tserkovnyak03,taniguchi15PRApplied}], however, in different systems 
compared with ours. 

% ================================================================================================================================================================================= %

There are even many other methods deriving the spin torque formula in the presence of the interface scattering [\onlinecite{slonczewski96,kovalev02,stiles02,barnas05,theodonis06}]. 
A spin torque formula including interface effect is often complex, as can be seen in the references mentioned above. 
In this work, we use the first approach mentioned above to calculate the coupling spin torque 
because it provides more accurate evaluation of the coupling spin torque. 
Before further discussing the spin torque formula, however, 
let us consider an approach to include the spin pumping in the present formalism. 

% ================================================================================================================================================================================= %

\subsection{Spin pumping}
\label{sec:Spin pumping}

The discussion in Sec. \ref{sec:Spin current in nonmagnetic connector} is valid when $\dot{\mathbf{m}}_{\ell}=\bm{0}$. 
On the other hand, when the magnetization dynamics is excited by the spin torque given by Eq. (\ref{eq:STT_1}), 
the spin pumping becomes another and unavoidable source of the spin current flowing in the nonmagnetic connector. 
In this section, including the effect of the spin pumping in the above formulas is discussed. 

The spin pumping is a phenomenon where a pure spin current is emitted from a ferromagnet to an adjacent metal as a result of the magnetization dynamics. 
The pumped spin current density at the F${}_{\ell}$/N${}^{\prime}$ is given by [\onlinecite{tserkovnyak02a}] 
\begin{equation}
  \mathbf{J}_{\rm s}^{{\rm pump}(\ell)}
  =
  \frac{\hbar}{4\pi S^{\prime}}
  g_{\rm r}^{\prime}
  \mathbf{m}_{\ell}
  \times
  \frac{d \mathbf{m}_{\ell}}{dt}.
  \label{eq:spin_pumping}
\end{equation}
In the presence of the spin pumping, 
the total spin current at the F${}_{\ell}$/N${}^{\prime}$ interface becomes 
$\mathbf{J}_{\rm s}^{{\rm F}_{\ell}/{\rm N}^{\prime}}=\mathbf{J}_{\rm s}^{{\rm pump}(\ell)}+\mathbf{J}_{\rm s}^{{\rm F}_{\ell}\to{\rm N}^{\prime}}$, i.e., 
\begin{equation}
\begin{split}
  \mathbf{J}_{\rm s}^{{\rm F}_{\ell}/{\rm N}^{\prime}}
  =&
  \frac{-1}{2\pi S^{\prime}}
  \left[
  \frac{(1-p_{\rm g}^{\prime\ 2})g^{\prime}}{2}
    \left(
      \mathbf{m}_{\ell}
      \cdot
      \delta
      \bm{\mu}_{\rm N^{\prime}}
    \right)
    \mathbf{m}_{\ell}
  \right.
\\
  &
  \left.
    +
    g_{\rm r}^{\prime}
    \mathbf{m}_{\ell}
    \times
    \left(
      \delta
      \bm{\mu}_{\rm N^{\prime}}
      \times
      \mathbf{m}_{\ell}
    \right)
  \right]
\\
  &+
  \frac{(1-p_{\rm g}^{\prime\ 2})g^{\prime}}{4\pi S^{\prime}}
  e \vartheta^{*}
  \lambda_{\rm F}
  E_{x}
  m_{\ell y}
  \mathbf{m}_{\ell}
  +
  \frac{\hbar}{4\pi S^{\prime}}
  g_{\rm r}^{\prime}
  \mathbf{m}_{\ell}
  \times
  \frac{d \mathbf{m}_{\ell}}{d t}. 
  \label{eq:spin_current_FN_appendix_2}
\end{split}
\end{equation}
Accordingly, Eq. (\ref{eq:spin_accumulation_N_p_appendix}) is modified as 
\begin{equation}
\begin{split}
  \delta
  \bm{\mu}_{\rm N^{\prime}}
  =
  \frac{2e^{2} \lambda_{\rm N^{\prime}}}{\hbar \sigma_{\rm N^{\prime}} \sinh(L/\lambda_{\rm N^{\prime}})}
  &
  \left[
    \mathbf{J}_{\rm s}^{{\rm F_{1}}/{\rm N^{\prime}}}
    \cosh
    \left(
      \frac{y-L}{\lambda_{\rm N^{\prime}}}
    \right)
  \right.
\\
  &
  \left.
    +
    \mathbf{J}_{\rm s}^{{\rm F_{2}}/{\rm N^{\prime}}}
    \cosh
    \left(
      \frac{y}{\lambda_{\rm N^{\prime}}}
    \right)
  \right],
  \label{eq:spin_accumulation_N_p_appendix_spin_pumping}
\end{split}
\end{equation}
i.e., $\mathbf{J}_{\rm s}^{{\rm F}_{\ell} \to {\rm N}^{\prime}}$ in Eq. (\ref{eq:spin_accumulation_N_p_appendix}) is replaced by 
the total spin current density $\mathbf{J}_{\rm s}^{{\rm F}_{\ell}/{\rm N}^{\prime}}$ including the spin pumping effect. 
Substituting Eq. (\ref{eq:spin_accumulation_N_p_appendix_spin_pumping}) into Eq. (\ref{eq:spin_current_FN_appendix_2}), 
the spin current at the F${}_{\ell}$/N${}^{\prime}$ interface, as well as the spin torque excited at the interface, can be calculated; 
see Sec. \ref{sec:Definition of coupling spin torque}. 

Before ending this section, we note that the spin pumping from the ferromagnet emits spin current 
not only to the nonmagnetic connector N${}^{\prime}$ but also to the bottom nonmagnet N. 
The spin torque due to the spin pumping into the nonmagnetic connector also excites a spin torque given by [\onlinecite{tserkovnyak02b}] 
\begin{equation}
  \mathbf{T}_{\ell}^{{\rm SP}}
  =
  \alpha^{\prime\prime}
  \mathbf{m}_{\ell}
  \times
  \frac{d \mathbf{m}_{\ell}}{dt},
\end{equation}
where $\alpha^{\prime\prime}$ is 
\begin{equation}
  \alpha^{\prime\prime}
  =
  \frac{\gamma \hbar}{4\pi MSd_{\rm F}}
  g_{\rm r}
  \left[
    1
    +
    \frac{g_{\rm r}}{g_{\rm N} \tanh(d_{\rm N}/\lambda_{\rm N})}
  \right]^{-1}.
  \label{eq:alpha_pp}
\end{equation}

% ================================================================================================================================================================================= %

\subsection{Definition of coupling spin torque}
\label{sec:Definition of coupling spin torque}

Substituting Eq. (\ref{eq:spin_accumulation_N_p_appendix_spin_pumping}) into Eq. (\ref{eq:spin_current_FN_appendix_2}), 
we find that the spin current at the F${}_{\ell}$/N${}^{\prime}$ interface is obtained by solving the following equations;
\begin{equation}
  \begin{pmatrix}
    \mathsf{D}^{(1)} & \mathsf{N}^{(1)} \\
    \mathsf{N}^{(2)} & \mathsf{D}^{(2)}
  \end{pmatrix}
  \begin{pmatrix}
    \mathbf{e}_{x}\cdot\mathbf{J}_{\rm s}^{{\rm F}_{1}/{\rm N}^{\prime}} \\
    \mathbf{e}_{y}\cdot\mathbf{J}_{\rm s}^{{\rm F}_{1}/{\rm N}^{\prime}} \\
    \mathbf{e}_{z}\cdot\mathbf{J}_{\rm s}^{{\rm F}_{1}/{\rm N}^{\prime}} \\
    \mathbf{e}_{x}\cdot\mathbf{J}_{\rm s}^{{\rm F}_{2}/{\rm N}^{\prime}} \\
    \mathbf{e}_{y}\cdot\mathbf{J}_{\rm s}^{{\rm F}_{2}/{\rm N}^{\prime}} \\
    \mathbf{e}_{z}\cdot\mathbf{J}_{\rm s}^{{\rm F}_{2}/{\rm N}^{\prime}}     
  \end{pmatrix}
  =
  \begin{pmatrix}
    \mathbf{e}_{x}\cdot\mathbf{J}_{\rm s}^{(1)} \\
    \mathbf{e}_{y}\cdot\mathbf{J}_{\rm s}^{(1)} \\
    \mathbf{e}_{z}\cdot\mathbf{J}_{\rm s}^{(1)} \\
    \mathbf{e}_{x}\cdot\mathbf{J}_{\rm s}^{(2)} \\
    \mathbf{e}_{y}\cdot\mathbf{J}_{\rm s}^{(2)} \\
    \mathbf{e}_{z}\cdot\mathbf{J}_{\rm s}^{(2)}     
  \end{pmatrix},
  \label{eq:current_equation}
\end{equation}
where $\mathbf{J}_{\rm s}^{(\ell)}=\mathbf{J}_{\rm s}^{{\rm SHE}(\ell)}+\mathbf{J}_{\rm s}^{{\rm pump}(\ell)}$ on the right hand side 
is the source term in Eq. (\ref{eq:spin_current_FN_appendix_2}), i.e., 
\begin{equation}
  \mathbf{J}_{\rm s}^{(\ell)}
  =
  \frac{(1-p_{\rm g}^{\prime\ 2})g^{\prime}}{4\pi S^{\prime}}
  e \vartheta^{*}
  \lambda_{\rm F}
  E_{x}
  m_{\ell y}
  \mathbf{m}_{\ell}
  +
  \frac{\hbar}{4\pi S^{\prime}}
  g_{\rm r}^{\prime}
  \mathbf{m}_{\ell}
  \times
  \frac{d \mathbf{m}_{\ell}}{d t}. 
  \label{eq:spin_current_source_term}
\end{equation}
The $(a,b)$ ($a,b=x,y,z$ or $1$,$2$,$3$) components of $3 \times 3$ matrices, $\mathsf{D}^{(\ell)}$ and $\mathsf{N}^{(\ell)}$, in Eq. (\ref{eq:current_equation}) are given by 
\begin{equation}
\begin{split}
  D_{ab}^{(\ell)}
  =&
  \delta_{ab}
  +
  \frac{(1-p_{g}^{\prime 2})g^{\prime}}{2g_{\rm N^{\prime}} \tanh(L/\lambda_{\rm N^{\prime}})}
  m_{\ell a}
  m_{\ell b}
\\
  &
  +
  \frac{g_{\rm r}^{\prime}}{g_{\rm N^{\prime}} \tanh(L/\lambda_{\rm N^{\prime}})}
  \left(
    \delta_{ab}
    -
    m_{\ell a}
    m_{\ell b}
  \right),
  \label{eq:matrix_D}
\end{split}
\end{equation}
\begin{equation}
\begin{split}
  N_{ab}^{(\ell)}
  =&
  \frac{(1-p_{g}^{\prime 2})g^{\prime}}{2g_{\rm N^{\prime}} \sinh(L/\lambda_{\rm N^{\prime}})}
  m_{\ell a}
  m_{\ell b}
\\
  &
  +
  \frac{g_{\rm r}^{\prime}}{g_{\rm N^{\prime}} \sinh(L/\lambda_{\rm N^{\prime}})}
  \left(
    \delta_{ab}
    -
    m_{\ell a}
    m_{\ell b}
  \right), 
  \label{eq:matrix_N}
\end{split}
\end{equation}
where $g_{\rm N^{\prime}}/S=h\sigma_{\rm N^{\prime}}/(2e^{2} \lambda_{\rm N^{\prime}})$. 
By solving Eq. (\ref{eq:current_equation}), the total spin current density $\mathbf{J}_{\rm s}^{{\rm F}_{\ell}/{\rm N^{\prime}}}$ at the F${}_{\ell}$/N${}^{\prime}$ is obtained. 
The spin torque excited at the F${}_{\ell}$/N${}^{\prime}$ interface is then calculated as 
\begin{equation}
  \mathbf{T}_{\ell}^{(2)}
  =
  \frac{\gamma}{Md_{\rm F}}
  \mathbf{m}_{\ell}
  \times
  \left(
    \mathbf{J}_{\rm s}^{{\rm F}_{\ell}/{\rm N^{\prime}}}
    \times
    \mathbf{m}_{\ell}
  \right).
  \label{eq:STT_2_def}
\end{equation}
As implied from Eq. (\ref{eq:current_equation}), 
the spin torque given by Eq. (\ref{eq:STT_2_def}) generally depends on the magnetizations of two STOs, $\mathbf{m}_{1}$ and $\mathbf{m}_{2}$. 
In other words, the spin torque given by Eq. (\ref{eq:STT_2_def}) acting on $\mathbf{m}_{\ell}$ depends on 
the magnetization in the other STO, $\mathbf{m}_{\ell^{\prime}}$ ($\ell,\ell^{\prime}=1,2$ and $\ell^{\prime} \neq \ell$). 
As a result, a coupled motion of the magnetizations is excited in the STOs. 
In Sec. \ref{sec:Coupled magnetization dynamics}, we investigate such coupled dynamics 
by solving the LLG equation with Eq. (\ref{eq:STT_2_def}). 

% ===================================================================================================================================================================================== %

At the end of this section, let us give some comments on Eq. (\ref{eq:STT_2_def}). 
According to the existence of two source terms of the spin current in the nonmagnetic connector shown in Eq. (\ref{eq:spin_current_source_term}), 
the torque given by Eq. (\ref{eq:STT_2_def}) can be decomposed into two contributions as 
\begin{equation}
  \mathbf{T}_{\ell}^{(2)}
  =
  \mathbf{T}_{\ell}^{(2){\rm SHE}}
  (\mathbf{m}_{1},\mathbf{m}_{2})
  +
  \mathbf{T}_{\ell}^{(2){\rm SP}}
  (\mathbf{m}_{1},\mathbf{m}_{2},\dot{\mathbf{m}}_{1},\dot{\mathbf{m}}_{2}).
  \label{eq:STT_2_decomp}
\end{equation}
Here, the first term, $\mathbf{T}_{\ell}^{(2){\rm SHE}}$, originates from 
the source term given by Eq. (\ref{eq:spin_current_SHE}), 
and depends on the magnetizations directions $\mathbf{m}_{1}$ and $\mathbf{m}_{2}$. 
On the other hand, the second term, $\mathbf{T}_{\ell}^{(2){\rm SP}}$, originates from 
the spin pumping given by Eq. (\ref{eq:spin_pumping}). 
We emphasize here that the torque $\mathbf{T}_{\ell}^{(2){\rm SHE}}$ is newly found in this work. 
On the other hand, the role of the spin pumping on coupled STOs was studied in Ref. [\onlinecite{taniguchi18PRB}] for a current-perpendicular-to-plane (CPP) structure. 
We should, however, note that the formalism developed in Ref. [\onlinecite{taniguchi18PRB}] is slightly different from the present work 
because Ref. [\onlinecite{taniguchi18PRB}] uses the second approach mentioned in Sec. \ref{sec:Spin current in nonmagnetic connector} to calculate the spin torque, 
i.e., the conservation of the spin current in the nonmagnetic connector is assumed. 

It may be useful for readers to show the explicit form of the coupling spin torque $\mathbf{T}_{\ell}^{(2){\rm SHE}}$ to apprehend its physical insight. 
The exact solution of $\mathbf{T}_{\ell}^{(2){\rm SHE}}$ is, however, complex; see Appendix \ref{sec:AppendixA}. 
Therefore, in the numerical simulation of the LLG equation discussed in Sec. \ref{sec:Coupled magnetization dynamics}, 
the coupling spin torque $\mathbf{T}_{\ell}^{(2){\rm SHE}}$ is calculated numerically. 
In Sec. \ref{sec:Role of coupling spin torque}, on the other hand, we derive an approximated analytical expression of $\mathbf{T}_{\ell}^{(2){\rm SHE}}$ 
to clarify the role of the torque on the coupled dynamics of the magnetization. 

% ================================================================================================================================================================================= %

%\begin{table}
%\begin{tabular}{|l|r|r|} \hline
%  & $\mathbf{T}_{\ell}^{(2){\rm SHE}}$ & $\mathbf{T}_{\ell}^{(2){\rm SP}}$ \\ \hline
%When is it fnite? & even when $\dot{\mathbf{m}}_{\ell}=\bm{0}$ & only when $\dot{\mathbf{m}}_{\ell} \neq \bm{0}$ \\ \hline
%Strength & $\propto J_{0}$ and independent of $\dot{\mathbf{m}}_{\ell}$ & $\propto \dot{\mathbf{m}}_{\ell}$ and independent of $J_{0}$ \\ \hline
%What kind of synchronization is excited? & antiphase (see Sec. ) & in-phase \cite{taniguchi18PRB} \\ \hline
%\end{tabular}
%\caption{}
%\label{sec:table1}
%\end{table}

% ================================================================================================================================================================================= %

% ===================================================================================================================================================================================== %

% ===================================================================================================================================================================================== %

\subsection{Relation to previous works}
\label{sec:Relation to previous works}

Before proceeding to further calculations, let us discuss the relation between the present and previous works. 
In 2017, Kudo and Morie proposed an array of STOs for practical application of pattern recognition based on spintronics technology [\onlinecite{kudo17}]. 
Each STO has a bottom electrode consisting of a nonmagnetic heavy metal driving the spin Hall effect. 
The system also has a common top electrode. 
Therefore, the structure is similar to the present model. 
They considered, however, supplying electric power from the top electrode, which injects electric currents into the STOs. 
There are two kinds of spin torques excited in their geometry; 
one arises from the spin Hall effect in the bottom nonmagnet, 
and the other originates from the electric current directly injected from the top electrode. 
The crucial point is that the total resistance of the system between the top and bottom electrodes depend on the magnetization directions of STOs due to the tunnel magnetoresistance effect. 
Therefore, the magnitude of the electric current injected from the top electrode to one STO depends on the magnetization directions of the other STOs. 
As a result, the magnetization dynamics in STOs are coupled [\onlinecite{kudo17}]. 
On the other hand, the present model demonstrates that a coupling is spontaneously induced without applying the electric current from the top electrode 
because spin currents naturally flow according to the gradient of the spin accumulations in the top electrode and due to the spin pumping. 
In this respect, this work can be partially regarded as a theoretical study elaborating the hidden mechanism of coupling in previously proposed STO arrays. 

The mutual and self-synchronization of the STOs with the spin Hall effect have been demonstrated both experimentally and theoretically. 
The coupling is driven by, for example, the spin wave [\onlinecite{awad17}] or the feedback of the oscillating electric current [\onlinecite{bhuktare17}]. 
A mutual synchronization by the electric current generated by the spin Hall magnetoresistance [\onlinecite{nakayama13,althammer13,chen13}] was also proposed in our previous work [\onlinecite{taniguchi17}]. 
The coupling mechanism in Ref. [\onlinecite{taniguchi17}] is a long-range interaction due to the conservation law of the electric current. 
The strength of the coupling is, however, weak because it is proportional to the third order of the spin Hall angle. 
On the other hand, the coupling mechanism proposed here comes purely from the spin degree of freedom, 
where the spin currents driven by the gradient of the spin accumulation and spin pumping lead to the synchronized dynamics of the magnetizations. 
The coupling strength of $\mathbf{T}_{\ell}^{(2){\rm SHE}}$ is proportional to the first order of the spin Hall angle, and therefore, is relatively strong, 
although the interaction length is restricted by the spin diffusion length. 

The present system is suitable to study the role of the coupling by spin current on the magnetization dynamics. 
Since the ferromagnets are connected by a nonmagnet, we can exclude a possibility of the coupling through spin wave [\onlinecite{kaka05,mancoff05}]. 
Also, since the dipole interaction decays according to the inverse cube detection law, 
we expect that the dipole coupling can be excluded by setting a sufficiently long distance, but shorter than the spin diffusion length, between STOs. 
One might also consider a structure where two STOs are placed onto a single nonmagnet, instead of the model shown in Fig. \ref{fig:fig1}(b). 
The same coupling mechanism appears even in such geometry. 
At the same time, however, a different mechanism of the coupling due to the spin Hall magnetoresistance [\onlinecite{taniguchi17}] will also appear 
when the STOs have a common bottom nonmagnet. 
The purpose of this work is to clarify the role of the coupling spin torque generated by spin current on the magnetization dynamics, 
and the structure having two STOs and one common bottom nonmagnet is therefore excluded. 
We also note that, although we consider the spin Hall geometry in this work as an example, 
a coupled dynamics of the magnetizations driven by a spin current is expected even in a CPP structure. 

It might be currently difficult to connect STOs within a short distance. 
An integration of STOs will be, however, an inevitable topic in the field of spintronics. 
In the integrated system, the STOs will be assembled in a distance shorter than the spin diffusion length. 
The coupling mechanism proposed in this work hence will be of great interest in such situation.

% ===================================================================================================================================================================================== %

% ===================================================================================================================================================================================== %

% ===================================================================================================================================================================================== %

\section{Coupled magnetization dynamics}
\label{sec:Coupled magnetization dynamics}

In this section, we study the magnetization dynamics in the presence of the coupling torque by solving the LLG equation numerically. 

% ===================================================================================================================================================================================== %

\subsection{Equation of motion}
\label{sec:Equation of motion}

The spin torques derived in Sec. \ref{sec:Spin torque formula from spin transport theory} lead to the excitation of the magnetization dynamics in the ferromagnets. 
It was experimentally shown that the spin torque $\mathbf{T}_{\ell}^{(1)}$, given by Eq. (\ref{eq:STT_1}), originated from the spin current injected from the bottom nonmagnet by 
the spin Hall effect can excite an self-oscillation of an in-plane magnetized ferromagnet [\onlinecite{liu12}]. 
Therefore, we assume that the magnetic field $\mathbf{H}_{\ell}$ consists of an in-plane anisotropy field $H_{\rm K}$ and the demagnetization field $4\pi M$ in the perpendicular direction as 
\begin{equation}
  \mathbf{H}_{\ell}
  =
  H_{\rm K}
  m_{\ell y}
  \mathbf{e}_{y}
  -
  4\pi M 
  m_{\ell z}
  \mathbf{e}_{z}.
  \label{eq:field}
\end{equation}
In addition, in the present system, the spin current flowing in the top electrode also provides the spin torque $\mathbf{T}_{\ell}^{(2)}$ given by Eq. (\ref{eq:STT_2_def}). 
Therefore, the LLG equation describing the magnetization dynamics in the F${}_{\ell}$ layer is given by 
\begin{equation}
\begin{split}
  \frac{d \mathbf{m}_{\ell}}{d t}
  =&
  -\gamma
  \mathbf{m}_{\ell}
  \times
  \mathbf{H}_{\ell}
  +
  \left(
    \alpha
    +
    \alpha^{\prime\prime}
  \right)
  \mathbf{m}_{\ell}
  \times
  \frac{d \mathbf{m}_{\ell}}{dt}
  +
  \mathbf{T}_{\ell}^{(1)}
  +
  \mathbf{T}_{\ell}^{(2)}, 
  \label{eq:LLG}
\end{split}
\end{equation}
where $\gamma$ and $\alpha$ are the gyromagnetic ratio and the intrinsic Gilbert damping constant, respectively, 
whereas $\alpha^{\prime\prime}$ is given by Eq. (\ref{eq:alpha_pp}). 
The intrinsic damping constant is the damping constant in the absence of the spin pumping, 
and is on the order of $10^{-3}-10^{-2}$ [\onlinecite{oogane06}]. 
As mentioned with regard to the explanation of Eq. (\ref{eq:STT_2_decomp}), 
the torque $\mathbf{T}_{\ell}^{(2)}$ is decomposed into two contributions. 
As a result, the LLG equation given by Eq. (\ref{eq:LLG}) can be rewritten in the form of 
\begin{equation}
  \mathsf{L}
  \begin{pmatrix}
    \dot{\mathbf{m}}_{1} \\
    \dot{\mathbf{m}}_{2} 
  \end{pmatrix}
  =
  \begin{pmatrix}
    -\gamma \mathbf{m}_{1} \times \mathbf{H}_{1} + \mathbf{T}_{1}^{(1)}+ \mathbf{T}_{1}^{(2){\rm SHE}} \\
    -\gamma \mathbf{m}_{2} \times \mathbf{H}_{2} + \mathbf{T}_{2}^{(1)}+ \mathbf{T}_{2}^{(2){\rm SHE}}
  \end{pmatrix}, 
  \label{eq:LLG_spin_pumping}
\end{equation}
where the effect of the torque $\mathbf{T}_{\ell}^{(2){\rm SP}}$ due to the spin pumping into the connector 
is included in a $6 \times 6$ matrix $\mathsf{L}$. 
The explicit forms of $\mathbf{T}_{\ell}^{(2){\rm SHE}}$ and $\mathsf{L}$ 
and their values are summarized in Appendix \ref{sec:AppendixA}. 
In the following sections, we will show the solutions of $\mathbf{m}_{\ell}$ by numerically calculating Eq. (\ref{eq:LLG_spin_pumping}). 
The material parameters used in the following discussion are derived from recent experiments on the spin Hall magnetoresistance in W/CoFeB metallic bilayer [\onlinecite{kim16}], 
where $\rho_{\rm F}=1/\sigma_{\rm F}=1.6$ k$\Omega$nm, $\beta=0.72$, $\lambda_{\rm F}=1.0$ nm, $\rho_{\rm N}=1/\sigma_{\rm N}=1.25$ k$\Omega$nm, $\lambda_{\rm N}=1.2$ nm, and $\vartheta=0.27$, 
whereas $d_{\rm F}=2$ nm and $d_{\rm N}=3$ nm. 
We assume that $r=0.25$ k$\Omega$nm${}^{2}$, $p_{g}=0.50$, and $g_{\rm r}/S=25$ nm${}^{-2}$. 
Also, we use $M=1500$ emu/c.c. [\onlinecite{kim16}], $H_{\rm K}=200$ Oe, and $\gamma=1.764 \times 10^{7}$ rad/(Oe s). 
The value of the intrinsic damping constant $\alpha$ is mentioned below. 
The detail of the numerical methods to calculate the LLG equation is summarized in Appendix \ref{sec:AppendixB}.

% ===================================================================================================================================================================================== %

The magnetization initially stays near the stable state of $\mathbf{m}_{\ell}=+\mathbf{e}_{y}$. 
We give different initial conditions to the magnetizations $\mathbf{m}_{1}$ and $\mathbf{m}_{2}$. 
Therefore, in the absence of the coupling due to the spin currents in the connector, two magnetizations oscillate independently with different phases. 
In the presence of the coupling spin torque, on the other hand, the dynamics of the magnetizations interact each other. 
As a result, the phase difference between the magnetizations will be stabilized and attains a certain value. 
The purpose of this section is to investigate such phase synchronization. 

% ===================================================================================================================================================================================== %

% ===================================================================================================================================================================================== %

\subsection{Coupled dynamics in STOs}
\label{sec:Coupled dynamics in STOs}

In this section, we study the coupled dynamics of the magnetizations by solving Eq. (\ref{eq:LLG_spin_pumping}) numerically. 

% ===================================================================================================================================================================================== %

% ===================================================================================================================================================================================== %

\begin{figure}%[p]
\centerline{\includegraphics[width=1.0\columnwidth]{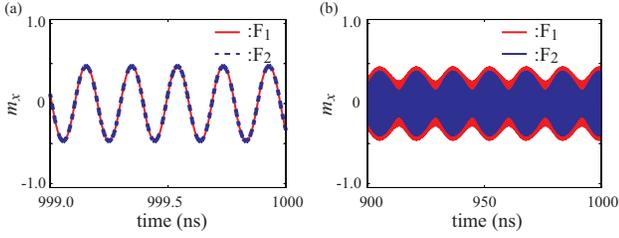}}%\vspace{-3.0ex}
\caption{
         Time evolutions of $m_{1x}$ (red) and $m_{2x}$ (blue) for (a) $J_{0}=44$ and (b) $48$ MA/cm${}^{2}$. 
         The intrinsic damping constant is $\alpha=0.005$. 
         Note that the time ranges of (a) and (b) are different. 
         \vspace{-3ex}}
\label{fig:fig2}
\end{figure}

% ===================================================================================================================================================================================== %

% ===================================================================================================================================================================================== %

First, let us assume that the damping constant $\alpha$ is relatively small, $\alpha=0.005$. 
In this case, we find that the magnetizations stay in the stable state $\mathbf{m}_{\ell}=+\mathbf{e}_{y}$ near the initial state for $J_{0}\lesssim 44$ MA/cm${}^{2}$ 
and switch their directions to the other stable state $\mathbf{m}_{\ell}=-\mathbf{e}_{y}$ for $J_{0} \gtrsim 60$ MA/cm${}^{2}$. 
Therefore, the oscillations of the magnetizations are excited when $44 \lesssim J_{0} \lesssim 60$ MA/cm${}^{2}$. 
Figure \ref{fig:fig2}(a) shows the oscillations of $m_{1x}$ (red solid line) and $m_{2x}$ (blue dashed line) in a steady state 
near the critical point $J_{0}=44$ MA/cm${}^{2}$. 
As shown, an in-phase synchronization of the magnetizations is excited. 
However, when the current magnitude is slightly increased to $J_{0}\simeq 48$ MA/cm${}^{2}$, the in-phase synchronization disappears. 
Instead, an oscillation accompanying a long-period beat appears, as can be seen in Fig. \ref{fig:fig2}(b), 
where we note that the time range of Fig. \ref{fig:fig2}(b) in the horizontal axis is different from that in Fig. \ref{fig:fig2}(a). 
In this case, a steady sate depends on the initial state of the magnetizations, and 
accordingly, the phase difference in this region is not well-defined. 

% ===================================================================================================================================================================================== %

% ===================================================================================================================================================================================== %

\begin{figure}%[p]
\centerline{\includegraphics[width=1.0\columnwidth]{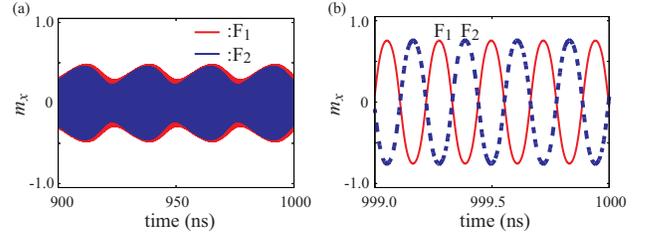}}%\vspace{-3.0ex}
\caption{
         Time evolutions of $m_{1x}$ (red) and $m_{2x}$ (blue) for (a) $J_{0}=72$ and (b) $82$ MA/cm${}^{2}$. 
         The intrinsic damping constant is $\alpha=0.010$. 
         Note that the time ranges of (a) and (b) are different. 
         \vspace{-3ex}}
\label{fig:fig3}
\end{figure}

% ===================================================================================================================================================================================== %

% ===================================================================================================================================================================================== %

Next, let us consider a relatively large damping case, $\alpha=0.010$. 
In this case, we observe the oscillations of the magnetizations in the current range of $72 \lesssim J_{0} \lesssim 82$ MA/cm${}^{2}$. 
Figure \ref{fig:fig3}(a) shows the oscillations of $m_{1x}$ and $m_{2x}$ near the critical point $J_{0}=72$ MA/cm${}^{2}$. 
Similar to Fig. \ref{fig:fig2}(b), the oscillation accompanying beat having a long period appears. 
When the current magnitude is increased to $J_{0}\simeq 82$ MA/cm${}^{2}$, however, the beat virtually disappears, 
and antiphase synchronization between STOs is excited, as can be seen in Fig. \ref{fig:fig3}(b).

% ===================================================================================================================================================================================== %

The in-phase synchronization is useful to enhance the emission power generated from an array of STOs, and thus, is useful for practical applications. 
On the other hand, the antiphase synchronization, or more generally an out-of-phase synchronization, 
has been found in several physical system such as Huygens pendulum clock [\onlinecite{pikovsky03,balanov09}], 
and therefore, is of great interest from the viewpoint of fundamental physics. 
One may consider that the antiphase synchronization is not preferable for practical applications. 
This is because the oscillating signals become totally zero or at least attenuate, 
and therefore, is not beneficial in terms of the enhancement of emission power from the devices. 
The out-of-phase synchronization, however, becomes of interest not only from the perspective of nonlinear science but also from 
for practical application viewpoint such as a phased array [\onlinecite{sun13}] and brain-inspired computing [\onlinecite{kudo17,vassilieva11}].

% ===================================================================================================================================================================================== %

% ===================================================================================================================================================================================== %

\subsection{Summary of this section and question to be addressed}
\label{sec:Summary of this section and question to be addressed}

The numerical simulations shown in Figs. \ref{fig:fig2} and \ref{fig:fig3} indicate that 
the steady state of the coupled magnetization dynamics depends on the damping constant and current magnitude. 
For small damping and small current region, the in-phase synchronization is stabilized. 
On the other hand, for a large damping case, the antiphase synchronization appears. 
In the intermediate region, the oscillation with a long-period beat is observed. 

It has been revealed theoretically in a CPP structure that the coupling between in-plane magnetized STO via spin pumping results in an in-phase synchronization [\onlinecite{taniguchi18PRB}]. 
Therefore, one might consider that the in-phase synchronization shown in Fig. \ref{fig:fig2}(a) is caused by the spin pumping. 
However, the origin of the antiphase synchronization in Fig. \ref{fig:fig3}(b), as well as that of the beat shown in Figs. \ref{fig:fig2}(b) and \ref{fig:fig3}(a), is unclear. 
We emphasize that there are two origins of the spin currents leading to the coupled motion of the magnetizations given by Eqs. (\ref{eq:spin_current_SHE}) and (\ref{eq:spin_pumping}). 
Therefore, to fully understand the results shown in Figs. \ref{fig:fig2} and \ref{fig:fig3}, 
it is necessary to reveal the role of the coupling spin torque $\mathbf{T}_{\ell}^{(2){\rm SHE}}$ originated from the source term of Eq. (\ref{eq:spin_current_SHE}). 
This analysis is developed in the next section.

% ===================================================================================================================================================================================== %

% ===================================================================================================================================================================================== %

\section{Role of coupling spin torque}
\label{sec:Role of coupling spin torque}

As shown in Sec. \ref{sec:Coupled magnetization dynamics}, 
the STOs show several kinds of the coupled dynamics, depending on the damping constant and current density. 
The purpose of this section is to clarify the physical insight of these results. 
Our recent work [\onlinecite{taniguchi18PRB}] already revealed that 
the coupling due to the spin pumping results in an in-phase synchronization between in-plane magnetized STOs, 
where the effect of the spin pumping in this study is described by the torque $\mathbf{T}_{\ell}^{(2){\rm SP}}$. 
On the other hand, the coupling spin torque $\mathbf{T}_{\ell}^{(2){\rm SHE}}$ is newly proposed in this study, 
and therefore, its role on the magnetization dynamics is not revealed yet. 
Therefore, in this section, we will concentrate on the magnetization dynamics in the presence of $\mathbf{T}_{\ell}^{(2){\rm SHE}}$ 
but neglecting the spin pumping effect. 
This approach clarifies the fact that the coupling spin torque $\mathbf{T}_{\ell}^{(2){\rm SHE}}$ prefers the antiphase synchronization between STOs, 
and provides a physical picture which is useful to understand the results shown in Figs. \ref{fig:fig2} and \ref{fig:fig3}. 

% ===================================================================================================================================================================================== %

\subsection{Approximated formula of coupling torque}
\label{sec:Approximated formula of coupling torque}

We remind the readers that the explicit form of the coupling spin torque $\mathbf{T}_{\ell}^{(2){\rm SHE}}$ is given as Eq. (\ref{eq:STT_2_revised}). 
However, the formula is complex and is not useful in the following analysis. 
Therefore, let us first derive another expression of $\mathbf{T}_{\ell}^{(2){\rm SHE}}$. 
Here, we use an approximation that the F${}_{\ell}$/N${}^{\prime}$ interface is transparent. 
In this case, the spin accumulation is continuous at the interface, contrary to the assumption used in Eq. (\ref{eq:spin_current_FN_appendix}), 
where the spin accumulation at the interface is discontinuous. 
When the interface is transparent, 
the solution of the spin accumulation in the N${}^{\prime}$ layer is given by 
\begin{equation}
\begin{split}
  \delta
  \bm{\mu}_{\rm N^{\prime}}(y)
  =
  \frac{1}{\sinh(L/\lambda_{\rm N^{\prime}})}
  &
  \left[
    \delta
    \bm{\mu}_{\rm F_{2}}(z=d_{\rm F})
    \sinh
    \left(
      \frac{y}{\lambda_{\rm N^{\prime}}}
    \right)
  \right.
\\
  &
  \left.
    -
    \delta
    \bm{\mu}_{\rm F_{1}}(z=d_{\rm F})
    \sinh
    \left(
      \frac{y-L}{\lambda_{\rm N^{\prime}}}
    \right)
  \right].
  \label{eq:spin_accumulation_mu_N_p}
\end{split}
\end{equation}
The spin current inside the N${}^{\prime}$ layer is given by 
\begin{equation}
  \mathbf{J}_{{\rm s}({\rm N}^{\prime})}
  =
  -\frac{\hbar \sigma_{{\rm N}^{\prime}}}{2e^{2}}
  \frac{\partial \delta\bm{\mu}_{{\rm N}^{\prime}}}{\partial y},
\end{equation}
When the spin diffusion length of the N${}^{\prime}$ layer is much longer than its length,  
we notice that 
\begin{equation}
\begin{split}
  \frac{\partial \delta\bm{\mu}_{\rm N^{\prime}}}{\partial y}
  &
  \simeq
  \frac{\delta\bm{\mu}_{\rm F_{2}}(z=d_{\rm F})-\delta\bm{\mu}_{\rm F_{1}}(z=d_{\rm F})}{L}. 
\end{split}
\end{equation}
Therefore, 
in the limit of $L/\lambda_{\rm N^{\prime}} \ll 1$, 
the spin current density flowing in the N${}^{\prime}$ layer, from the F${}_{\ell}$ to F${}_{\ell^{\prime}}$ layer [$(\ell,\ell^{\prime})=(1,2)$ or (2,1)], can be approximated as, 
\begin{equation}
  \mathbf{J}_{\rm s}^{{\rm F}_{\ell} \to {\rm F}_{\ell^{\prime}}}
  \simeq
  \frac{\hbar \sigma_{{\rm N}^{\prime}}}{2e^{2}L}
  \left[
    \delta
    \bm{\mu}_{{\rm F}_{\ell}}(z=d_{\rm F})
    -
    \delta
    \bm{\mu}_{{\rm F}_{\ell^{\prime}}}(z=d_{\rm F})
  \right].
  \label{eq:spin_current_coupling} 
\end{equation}
The emission of the spin current at the top interface leads to the spin torque given by 
\begin{equation}
\begin{split}
  \tilde{\mathbf{T}}_{\ell}^{(2){\rm SHE}}
  &=
  \frac{\gamma}{Md_{\rm F}}
  \mathbf{m}_{\ell}
  \times
  \left(
    \mathbf{J}_{\rm s}^{{\rm F}_{\ell} \to {\rm F}_{\ell^{\prime}}}
    \times
    \mathbf{m}_{\ell}
  \right)
\\
  &=
  -\frac{\gamma \hbar \tilde{\vartheta} J_{0}}{2eMd_{\rm F}}
  m_{\ell^{\prime}y}
  \mathbf{m}_{\ell}
  \times
  \left(
    \mathbf{m}_{\ell^{\prime}}
    \times
    \mathbf{m}_{\ell}
  \right),
  \label{eq:STT_2}
\end{split}
\end{equation}
where $\tilde{\vartheta}$ is defined as 
\begin{equation}
\begin{split}
  \tilde{\vartheta}
  &=
  \vartheta^{*}
  \frac{\sigma_{\rm N^{\prime}} \lambda_{\rm F}}{\sigma_{\rm N} L}
\\
  &=
  \vartheta
  \frac{\sigma_{\rm N^{\prime}} g^{*} \lambda_{\rm F} \tanh[d_{\rm N}/(2 \lambda_{\rm N})]}{(1-\beta^{2}) \sigma_{\rm F} g_{\rm N} L \sinh(d_{\rm F}/\lambda_{\rm F})}. 
  \label{eq:coupling_constant}
\end{split}
\end{equation}
The value of the coupling constant $\tilde{\vartheta}$ is $\tilde{\vartheta}/\vartheta\simeq 0.1$ for the present parameter. 
Equation (\ref{eq:STT_2}) is the approximated formula of $\mathbf{T}_{\ell}^{(2){\rm SHE}}$ given by Eq. (\ref{eq:STT_2_revised}). 
We emphasize that using Eq. (\ref{eq:STT_2}), instead of Eq. (\ref{eq:STT_2_revised}), does not change the qualitative picture of magnetization dynamics; see also Appendix \ref{sec:AppendixA}. 
It should also be noted that Eq. (\ref{eq:STT_2}) is useful to develop an analytical theory of the phase synchronization 
because of its simplified form; see Sec. \ref{sec:Analytical theory} below. 

% ===================================================================================================================================================================================== %

\subsection{Scaling currents}
\label{sec:Scaling currents}

In the absence of the coupling spin torque, the spin Hall effect excites 
the self-oscillation of the magnetization around the in-plane easy ($y$) axis 
when the magnitude of the current density is in the range of $J_{\rm c} < |J_{0}| < J^{*}$, where
\begin{equation}
  J_{\rm c}
  =
  \frac{2 \alpha eMd_{\rm F}}{\hbar \vartheta_{\rm R}}
  \left(
    H_{\rm K}
    +
    2\pi M
  \right), 
  \label{eq:J_c}
\end{equation}
\begin{equation}
  J^{*}
  =
  \frac{4 \alpha eMd_{\rm F}}{\pi \hbar \vartheta_{\rm R}}
  \sqrt{4\pi M (H_{\rm K}+4\pi M)}. 
  \label{eq:J_star}
\end{equation}
The current $J_{\rm c}$, called critical current density [\onlinecite{grollier03}], is the minimum current necessary to destabilize the magnetization staying near the easy axis 
and excites self-oscillation. 
On the other hand, the current $J^{*}$ is the switching current density to reverse the magnetization direction between two stable states, $\mathbf{m}_{\ell}=\pm\mathbf{e}_{y}$. 
The magnetization oscillates around the positive (negative) direction of the $y$ axis 
when $\vartheta_{\rm R} J_{0}$ is positive (negative). 
%The values of $J_{\rm c}$ and $J^{*}$ given by Eqs. (\ref{eq:J_c}) and (\ref{eq:J_star}) are $26$ and $33$ MA/cm${}^{2}$, respectively. 
We note that the currents $J_{\rm c}$ and $J^{*}$ determining the oscillation region of the magnetization are slightly affected 
by the coupling spin torque, as will be mentioned in the next section. 

% ===================================================================================================================================================================================== %

\subsection{Numerical simulation}
\label{sec:Numerical simulation}

In this section, we solve the LLG equation in the presence of Eq. (\ref{eq:STT_2}). 
Since the purpose of this section is to clarify the role of Eq. (\ref{eq:STT_2}), we have neglected the spin pumping effect in this section. 
Therefore, the LLG equation is given by 
\begin{equation}
\begin{split}
  \frac{d \mathbf{m}_{\ell}}{d t}
  =&
  -\gamma
  \mathbf{m}_{\ell}
  \times
  \mathbf{H}_{\ell}
  +
  \alpha
  \mathbf{m}_{\ell}
  \times
  \frac{d \mathbf{m}_{\ell}}{dt}
  +
  \mathbf{T}_{\ell}^{(1)}
  +
  \tilde{\mathbf{T}}_{\ell}^{(2){\rm SHE}}, 
  \label{eq:LLG}
\end{split}
\end{equation}
For the damping constant, we use $\alpha=0.005$. 
The values of $J_{\rm c}$ and $J^{*}$ are $26$ and $33$ MA/cm${}^{2}$, respectively. 

% ===================================================================================================================================================================================== %

% ===================================================================================================================================================================================== %

% ===================================================================================================================================================================================== %

\begin{figure}%[p]
\centerline{\includegraphics[width=1.0\columnwidth]{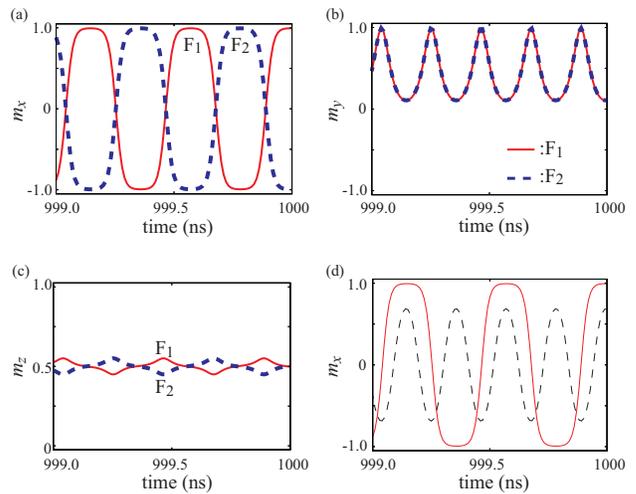}}%\vspace{-3.0ex}
\caption{
         The synchronized motions of the $x$, $y$, and $z$ components of $\mathbf{m}_{1}$ (red solid line) 
         and $\mathbf{m}_{2}$ (blue dotted line) at $J_{0}=28$ MA/cm${}^{2}$ are shown in (a), (b), and (c), respectively. 
          The coupled and free-running (uncoupled) oscillations of $m_{1x}$ at $J_{0}=28$ MA/cm${}^{2}$ are shown in (d) by the red solid and black dashed lines, respectively. 
         \vspace{-3ex}}
\label{fig:fig4}
\end{figure}

% ===================================================================================================================================================================================== %

% ===================================================================================================================================================================================== %

% ===================================================================================================================================================================================== %

Figures \ref{fig:fig4}(a), \ref{fig:fig4}(b), and \ref{fig:fig4}(c) respectively show the examples of 
the oscillations of the $x$, $y$, and $z$ components of $\mathbf{m}_{1}$ (red solid line) and $\mathbf{m}_{2}$ (blue dotted line) for $J_{0}=28$ MA/cm${}^{2}$. 
The results indicate that an antiphase synchronization of two magnetizations is excited by the coupling torque given by Eq. (\ref{eq:STT_2}). 
Notice that the oscillation trajectory is suppressed along the $z$ direction due to the large demagnetization field. 
The oscillation trajectory is well described by the elliptic functions, as discussed in Appendix \ref{sec:AppendixC}. 
We will revisit this point in the next section to investigate the stable phase-difference by an analytical calculation of the LLG equation. 
Although the results shown in Figs. \ref{fig:fig4}(a)-\ref{fig:fig4}(c) are examples for one set of parameters and initial conditions, 
we confirmed that the antiphase synchronization of two magnetizations is achieved for a wide range of the coupling constant $\tilde{\vartheta}$, current density $J_{0}$, and initial conditions. 

% ===================================================================================================================================================================================== %

% ===================================================================================================================================================================================== %

% ===================================================================================================================================================================================== %

\begin{figure}%[p]
\centerline{\includegraphics[width=1.0\columnwidth]{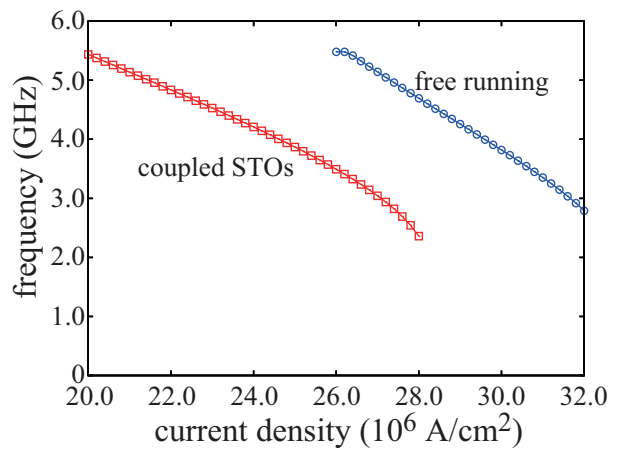}}%\vspace{-3.0ex}
\caption{
         Dependences of the oscillation frequencies of the free-running (blue circle) and coupled (red square) STOs on the current density $J_{0}$. 
         \vspace{-3ex}}
\label{fig:fig5}
\end{figure}

% ===================================================================================================================================================================================== %

% ===================================================================================================================================================================================== %

% ===================================================================================================================================================================================== %

Figure \ref{fig:fig4}(d) compares the oscillations of $m_{1x}$ for $J_{0}=28$ MA/cm${}^{2}$ for the coupled and uncoupled (free-running) STOs by the red solid and black dashed lines, respectively. 
We notice that the oscillation frequency of the coupled oscillator, 2.4 GHz, is lower than the free-running frequency, 4.7 GHz. 
Figure \ref{fig:fig5} summarizes the dependences of the oscillation frequencies of the coupled and free-running STOs on a wide range of $J_{0}$. 
The frequency dependence on current shifts to the low current region due to the coupling. 
This result can be explained as follows. 
The current range for the self-oscillation is determined by two characteristic current densities $J_{\rm c}$ and $J^{*}$ given by Eqs. (\ref{eq:J_c}) and (\ref{eq:J_star}), as mentioned earlier. 
We find that $J_{\rm c}$ and $J^{*}$ for coupled two identical STOs should instead be substituted to (see also Appendix \ref{sec:AppendixD}) 
\begin{equation}
  J_{\rm c}
  \to
  \frac{J_{\rm c}}{1+2(\tilde{\vartheta}/\vartheta_{\rm R})},
  \label{eq:J_c_coupled}
\end{equation}
\begin{equation}
  J^{*}
  \to
  \frac{J^{*}}{1+(\tilde{\vartheta}/\vartheta_{\rm R})[4\pi M/(H_{\rm K}+4\pi M)]}.
  \label{eq:J_star_coupled}
\end{equation}
Since the factors $2(\tilde{\vartheta}/\vartheta_{\rm R})$ and $(\tilde{\vartheta}/\vartheta_{\rm R})[4\pi M/(H_{\rm K}+4\pi M)]$ are positive, 
the values of $J_{\rm c}$ and $J^{*}$ for two coupled STOs are smaller than those without coupling. 
The values of $J_{\rm c}$ and $J^{*}$ for two coupled STOs are $20$ and $28$ MA/cm${}^{2}$, respectively. 
Therefore, the curve that represents the relation between the current density and frequency shifts to a low current region due to the coupling. 

% ===================================================================================================================================================================================== %

\subsection{Analytical theory}
\label{sec:Analytical theory}

The numerical simulation indicates that the coupling spin torque given by Eq. (\ref{eq:STT_2}) prefers antiphase coupling of the magnetizations. 
This result implies that the coupling torque in the present model acts as a repulsive force of the phase in the oscillators. 
We notice that this conclusion can also be explained analytically. 

% ===================================================================================================================================================================================== %

A standard approach in the field of nonlinear science to clarify the stable phase difference between the coupled oscillators is 
to reduce the equation of motion to the Kuramoto model [\onlinecite{kuramoto03,strogatz01,pikovsky03,stankovski17}]. 
The Kuramoto model argues that the oscillation properties of any kind of oscillator are characterized by the phase $\psi$ defined from the oscillation period. 
This is based on the assumption that any kind of the oscillation trajectory can be transformed into the trigonometric functions after a proper transformation of the coordinate called phase reduction [\onlinecite{stankovski17}]. 
The phase in the proper coordinate always satisfy the relation $\psi=2\pi ft$, where $f$ is a constant oscillation frequency. 
Solving the equation of motion of the phase difference near the fixed points, the stability of the phase difference can be investigated. 
Therefore, we are naturally motivated to derive the Kuramoto model in the present system. 

On the other hand, in the experiments using STOs, 
the word "phase" has often been used to describe the oscillation of the electric power generated from the STO [\onlinecite{tsunegi16}], 
which directly reflects the magnetization oscillation in the real space. 
When the oscillation of the magnetization is described by the trigonometric functions with the frequency $f$, 
the phase $\phi$ measured in the experiments is simply given by $\phi=2\pi ft$. 
In general, however, the magnetization oscillation cannot be described exactly by the trigonometric function. 
For example, the trajectory of the self-oscillation for the in-plane magnetized STO is described by the elliptic function, 
as discussed in the last section and Appendix \ref{sec:AppendixC}. 
The oscillation trajectories in the other types of an STO are, except special cases [\onlinecite{silva10}], 
also not described by the trigonometric function 
due to, for example, symmetry breaking by an external magnetic field [\onlinecite{taniguchi15}] or angular dependence of spin torque [\onlinecite{taniguchi17PRB}]. 
Breaking a symmetry is often necessary in STO devices for an electrical detection of the oscillating signal 
because, in a highly symmetrical system, no electrical signal is obtained 
through giant magnetoresistance or tunnel magnetoresistance effect, 
even if a self-oscillation of the magnetization is excited. 
In these cases, the phase defined from the experiments is, strictly speaking, different from the phase in the Kuramoto model. 
It is difficult and/or complicated to apply the phase reduction or similar approximations to the LLG equation analytically [\onlinecite{slavin08,kudo10,li10,li11}]. 
Moreover, it has been shown that an extension of the Kuramoto model to the nonlinear region is necessary 
to describe the self-oscillation in STOs [\onlinecite{slavin09}], 
where the nonlinearity means that the oscillation frequency, as well as the phase, strongly couples to the oscillation amplitude. 

We can, nevertheless, show that the coupling spin torque in the present model acts as repulsive force between the STOs 
by deriving the Kuramoto model from the LLG equation. 
This approach will also be useful to discuss the frequency locking in the next section. 
When the amplitude of the self-oscillation is small, we can apply the linear approximation to the LLG equation, 
where the oscillation trajectory is well approximated by the trigonometric function. 
We note that a fixed point for the in-plane magnetized STO is $|\mathbf{m}_{\ell}|=\mathbf{e}_{y}$. 
In this case, it is convenient to introduce the zenith and azimuth angles $(\theta_{\ell},\varphi_{\ell})$ in a spherical coordinate as 
$\mathbf{m}_{\ell}=(\sin\theta_{\ell}\sin\varphi_{\ell},\cos\theta_{\ell},\sin\vartheta_{\ell}\cos\varphi_{\ell})$, 
although this definition is different from the conventional definitions of the zenith and azimuth angles. 
The reason why we use this modified relation between $(\theta_{\ell},\varphi_{\ell})$ and $\mathbf{m}_{\ell}$ is as follows. 
The assumption of the small amplitude oscillation means that $|\mathbf{m}_{\ell}| \simeq \mathbf{e}_{y}$, and therefore, $\theta_{\ell} \to 0,\pi$. 
In this limit, the oscillation of the magnetization around the $y$ axis is approximately described by the trigonometric function, 
and $\varphi_{\ell}$ can be directly regarded as the phase of the Kuramoto model; see Appendix \ref{sec:AppendixC}. 
As a result, we can determine the stable phase difference between STOs by using the Kuramoto model [\onlinecite{strogatz01}]. 
In the limit of $\theta_{\ell} \to 0,\pi$, the LLG equation for $\varphi_{\ell}$ becomes 
\begin{equation}
\begin{split}
  \frac{d \varphi_{\ell}}{dt}
  \simeq&
  \pm
  \gamma
  \left(
    H_{\rm K}
    +
    4 \pi M 
    \cos^{2}\varphi_{\ell}
  \right)
\\
  &+
  \gamma 
  H_{\rm s2}
  m_{\ell^{\prime}y}
  \sin
  \left(
    \varphi_{\ell}
    -
    \varphi_{\ell^{\prime}}
  \right),
  \label{eq:Kuramoto_model}
\end{split}
\end{equation}
where we introduce the notation 
\begin{equation}
  H_{\rm s2}
  =
  \frac{\hbar \tilde{\vartheta} J_{0}}{2eMd_{\rm F}}. 
  \label{eq:H_s2}
\end{equation}
The double sign $\pm$ means the upper for $\theta_{\ell} \to 0$ and the lower for $\theta_{\ell} \to \pi$. 
We neglect the spin torque $\mathbf{T}_{\ell}^{(1)}$ and the damping torque 
because these torques cancel each other to sustain the self-oscillation. 
We note that $m_{\ell^{\prime}y}$ in Eq. (\ref{eq:Kuramoto_model}) should be replaced by $+1(-1)$ 
when we focus on the small amplitude self-oscillation around the positive (negative) $y$ axis. 
As mentioned earlier in this study, the positive (negative) spin current $\vartheta_{\rm R}J_{0}$ excites 
the self-oscillation around the positive (negative) direction of the $y$ axis. 
In addition, the parameter $\tilde{\vartheta}$ has the same sign with $\vartheta_{\rm R}$. 
Therefore, we can replace $H_{\rm s2}m_{\ell^{\prime}y}$ in Eq. (\ref{eq:Kuramoto_model}) with $|H_{\rm s2}|$ in the present approximation. 

Let us define the phase difference between two STOs as $\Delta\varphi=\varphi_{1}-\varphi_{2}$. 
According to Eq. (\ref{eq:Kuramoto_model}), the phase difference near the in-phase ($\Delta\varphi=0$) or antiphase ($\Delta\varphi=\pi$) state obeys 
\begin{equation}
  \frac{d \Delta\varphi}{dt}
  =
  2 \gamma 
  |H_{\rm s2}|
  \sin
  \Delta
  \varphi. 
  \label{eq:LLG_phase_difference}
\end{equation}
%Equation (\ref{eq:LLG_phase_difference}) is known as a simplified Kuramoto model \cite{strogatz01}. 
Let us imagine that the phase difference slightly shifts from a fixed point corresponding to the in-phase state as $\Delta\varphi=0+\epsilon$ ($|\epsilon| \ll 1$). 
The shift $\epsilon$ obeys 
\begin{equation}
  \frac{d \epsilon}{dt}
  =
  2 \gamma 
  |H_{\rm s2}|
  \epsilon. 
\end{equation}
The solution of this equation is given by $\epsilon=C_{1}e^{2\gamma |H_{\rm s2}|t}$, where $C_{1}$ is the integral constant. 
This solution means that the shift from the in-phase state increases with time, indicating that the in-phase state is an unstable fixed point. 
On the other hand, a shift $\epsilon$ of the phase difference from the other fixed point corresponding to the antiphase state, defined as $\Delta\varphi=\pi+\epsilon$, obeys 
\begin{equation}
  \frac{d \epsilon}{dt}
  =
  -2 \gamma 
  |H_{\rm s2}|
  \epsilon. 
  \label{eq:LLG_phase_difference_antiphase}
\end{equation}
The solution $\epsilon=C_{2}e^{-2\gamma |H_{\rm s2}|t}$ decreases with time, indicating that the antiphase state is the stable fixed point. 
Therefore, the coupling spin torque given by Eq. (\ref{eq:STT_2}) leads to the antiphase synchronization of the magnetizations. 
As mentioned above, the calculation in this section is valid only for a small amplitude oscillation. 
The result of the numerical simulation, however, indicates that the analysis explained so far can also be applied to large amplitude oscillation. 

% ===================================================================================================================================================================================== %

% ===================================================================================================================================================================================== %

\subsection{Summary of this section and answer to the question in the last section}
\label{sec:Summary of this section and answer to the question in the last section}

The results shown in Secs. \ref{sec:Numerical simulation} and \ref{sec:Analytical theory} indicate that 
the coupling spin torque given by Eq. (\ref{eq:STT_2}) results in the antiphase synchronization between STOs. 
This conclusion explains the results shown in Figs. \ref{fig:fig2} and \ref{fig:fig3} in Sec. \ref{sec:Coupled dynamics in STOs} as follows. 

At first, we should remind the readers that the LLG equation used in Sec. \ref{sec:Coupled dynamics in STOs} includes two coupling mechanisms between STOs, 
where one originates from the spin accumulation whereas the other comes from the spin pumping. 
As implied in Eqs. (\ref{eq:spin_current_SHE}) and (\ref{eq:spin_pumping}), 
the strength of the coupling spin torque $\mathbf{T}_{\ell}^{(2){\rm SHE}}$ due to the spin accumulation is proportional to the current density $J_{0}$ 
whereas that of the spin pumping $\mathbf{T}_{\ell}^{(2){\rm SP}}$ is proportional to the oscillation frequency of the magnetization, $f \sim |\dot{\mathbf{m}}|/(2\pi)$. 
In addition, the oscillation frequency of the in-plane magnetized ferromagnet decreases with increasing the current density, as shown in Fig. \ref{fig:fig5}. 
These facts indicate that the coupling due to the spin pumping is dominated in a relatively low current region, 
whereas that originated from the spin accumulation becomes large in a relatively high current region. 
As clarified in Ref. [\onlinecite{taniguchi18PRB}], the spin pumping prefers the in-phase synchronization between the in-plane magnetized STOs. 
Therefore, the in-phase synchronization is found in Fig. \ref{fig:fig2}(a). 
When the current magnitude is increased, however, the spin torque due to the spin accumulation becomes a dominant contribution to the coupled motion of the magnetizations. 
Since this coupling torque prefers the antiphase synchronization, as clarified in Secs. \ref{sec:Numerical simulation} and \ref{sec:Analytical theory}, 
a beat of the oscillations appears in the intermediate current region shown in Figs. \ref{fig:fig2}(b) and \ref{fig:fig3}(a) 
as a result of the competition between two coupling mechanisms. 
When the current becomes sufficiently large, the antiphase synchronization is stabilized by the coupling torque originated from the spin accumulation, as can be seen in Fig. \ref{fig:fig3}(c).

% ===================================================================================================================================================================================== %

% ===================================================================================================================================================================================== %

\begin{figure}%[p]
\centerline{\includegraphics[width=1.0\columnwidth]{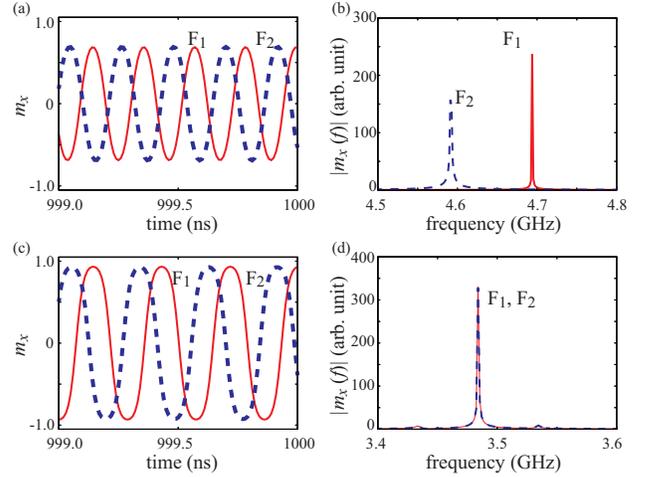}}%\vspace{-3.0ex}
\caption{
         (a) The oscillations of $m_{\ell x}$ $(\ell=1,2)$ and (b) their Fourier transformations, $|m_{\ell x}(f)|$, in the absence of the coupling. 
             The red solid and blue dotted lines correspond to the F${}_{1}$ and F${}_{2}$ layers, respectively. 
         (c) The oscillations of $m_{\ell x}$ and (d) their Fourier transformations for the coupled STOs. 
         \vspace{-3ex}}
\label{fig:fig6}
\end{figure}

% ===================================================================================================================================================================================== %

% ===================================================================================================================================================================================== %

% ===================================================================================================================================================================================== %

\subsection{Frequency locking}
\label{sec:Frequency locking}

In the above sections, we assume that two STOs have identical parameters. 
An interesting characteristic of synchronization is, on the other hand, the frequency locking. 
A typical value of the locking range of the frequency between STOs in the previous works is on the order of 10-100 MHz [\onlinecite{kaka05,mancoff05,rippard05}]. 
At the end of this section, 
let us briefly study the possibility of a frequency locking for two STOs having different free-running frequencies by the coupling torque given by Eq. (\ref{eq:STT_2}) 
because it provides another viewpoint to catch the strength of the coupling. 

The range of the frequency locking is determined by the strength of the coupling spin torque. 
Although it is difficult to derive an exact analytical expression of the locking range, 
Eq. (\ref{eq:LLG_phase_difference}) implies that the locking range is roughly given by 
\begin{equation}
  \delta f_{\rm lock}
  \sim
  \frac{2 \gamma |H_{\rm s2}|}{2\pi}
  =
  1.67 \times J_{0}\ 
  {\rm Hz/(A/cm{}^{2})}. 
  \label{eq:locking_range}
\end{equation}
The value of Eq. (\ref{eq:locking_range}) for the present system is about 40 MHz. 
%for the current density $J_{0}$ to excite the self-oscillation. 
In addition, the nonlinearity is expected to increase the locking range [\onlinecite{slavin09}]. 

% ===================================================================================================================================================================================== %

% ===================================================================================================================================================================================== %

% ===================================================================================================================================================================================== %

\begin{figure}%[p]
\centerline{\includegraphics[width=1.0\columnwidth]{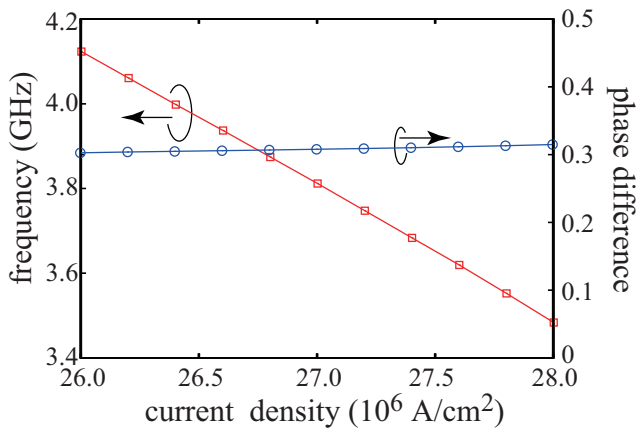}}%\vspace{-3.0ex}
\caption{
         Dependences of the oscillation frequencies (red square) and phase difference (blue circle) of the coupled STOs on the current density $J_{0}$. 
         \vspace{-3ex}}
\label{fig:fig7}
\end{figure}

% ===================================================================================================================================================================================== %

% ===================================================================================================================================================================================== %

% ===================================================================================================================================================================================== %

In the following, the possibility of the frequency locking is confirmed using numerical simulations. 
The value of $H_{\rm K}$ in the F${}_{2}$ layer is changed to 192 Oe to make the difference of the oscillation frequencies of STOs to about 100 MHz, 
which is the maximum value of the locking range we obtained from the numerical simulation. 
In Fig. \ref{fig:fig6}(a), we show the free-running oscillations of $m_{1x}$ and $m_{2x}$ by red solid and blue dotted lines, respectively, 
where the coupling spin torque is set to zero. 
As mentioned, the frequency difference between $m_{1x}$ and $m_{2x}$ is about 100 MHz, 
as can be seen from their Fourier transformations in Fig. \ref{fig:fig6}(b). 
On the other hand, in the presence of the coupling spin torque, 
the frequency locking is observed, as schematically shown in Figs. \ref{fig:fig6}(c) and \ref{fig:fig6}(d), 
where the oscillations of $m_{\ell x}$ ($\ell=1,2$) and their Fourier transformations are shown. 
Figure \ref{fig:fig7} shows the current dependences of the oscillation frequencies (red square) and phase difference (blue circle) of the coupled STOs. 
The in-phase and antiphase synchronizations correspond to the phase difference of 0 and 0.5, respectively (see also Appendix \ref{sec:AppendixB}). 
The frequency locking occurs for the current of $J_{0} \ge 26.0$ MA/cm${}^{2}$. 
We note that the locked frequency is different from the free-running frequencies of the STOs or their average,  
and the phase difference is different from the antiphase due to the frequency difference, as expected for the nonlinear oscillator [\onlinecite{slavin09}].  
The numerical simulation indicates that the phase difference in the locked state is almost constant.

\section{Summary}
\label{sec:Summary}

In conclusion, a theoretical framework for the spin-current driven synchronization in spin torque oscillators in the spin Hall geometry was proposed. 
The spin current generated from the spin Hall effect excites the self-oscillation of the magnetization 
and simultaneously creates the spin accumulation in the oscillator. 
The magnitude and direction of the spin accumulation in the ferromagnet depend on the magnetization direction of the oscillator. 
Then, by connecting the top surfaces of the oscillators with a nonmagnet having a long spin diffusion length, 
a spin current spontaneously flows between the oscillators, according to the gradient of the spin accumulation. 
The spin current excites an additional spin torque acting on the magnetization. 
As a result, the self-oscillations in the oscillators are naturally coupled. 
The coupling mechanism comes purely from the spin degree of freedom, contrary to the previous proposals based on the electric and/or magnetic interactions. 
Both the numerical simulation and analytical theory show that the coupling torque acts as a repulsive force, 
and therefore, the antiphase synchronization of the self-oscillation is preferred by this coupling mechanism. 
These conclusions are obtained by deriving the theoretical formulas of the coupling spin torque from the spin transport theory, 
and by solving the equation of motion of the magnetizations with the coupling spin torque both numerically and analytically. 
In the self-oscillations state, however, the spin pumping becomes another source of the coupling because 
the spin current generated by the spin pumping also flows in the nonmagnetic connector. 
When the spin pumping is taken into account, a competition between the coupling spin torque and spin pumping appears 
because the spin pumping prefers an in-phase synchronization. 
As a result, the in-phase synchronization appears in a relatively low current region whereas the antiphase synchronization appears in a relatively high current region. 
%It is also shown that the present coupling mechanism has a possibility to lock the frequency of the oscillators having the difference of the free-running frequencies about 100 MHz. 

% ===================================================================================================================================================================================== %

\section*{Acknowledgement}

The author is thankful to Yoji Kawamura, Kiwamu Kudo, Hitoshi Kubota, Sumito Tsunegi, Takehiko Yorozu, Hidekazu Saito, Takahiro Chiba, Yasuhiro Utsumi, and Masamitsu Hayashi for valuable discussions. 
The author is also grateful to Satoshi Iba, Aurelie Spiesser, Hiroki Maehara, and Ai Emura for their support and encouragement. 
This work was supported by JSPS KAKENHI Grant-in-Aid for Young Scientists (B) 16K17486.

% ================================================================================================================================================================================= %

\appendix

% ================================================================================================================================================================================= %

% ===================================================================================================================================================================================== %

\section{Explicit forms of $\mathbf{T}_{\ell}^{(2){\rm SHE}}$ and $\mathsf{L}$ and validity of $\tilde{\mathbf{T}}_{\ell}^{(2){\rm SHE}}$}
\label{sec:AppendixA}

In this Appendix, we show the explicit forms of $\mathbf{T}_{\ell}^{(2){\rm SHE}}$ and $\mathsf{L}$ in Eq. (\ref{eq:LLG_spin_pumping}). 
The validity of the approximated formula of $\mathbf{T}_{\ell}^{(2){\rm SHE}}$, $\tilde{\mathbf{T}}_{\ell}^{(2){\rm SHE}}$ given by Eq. (\ref{eq:STT_2}), is also discussed. 

% ===================================================================================================================================================================================== %

\subsection{Definitions of $\mathbf{T}_{\ell}^{(2){\rm SHE}}$ and $\mathsf{L}$}

To make the notation simple, 
we define a $6 \times 6$ matrix $\mathsf{M}$ from Eq. (\ref{eq:current_equation}) as 
\begin{equation}
  \mathsf{M}
  =
  \begin{pmatrix}
    \mathsf{D}^{(1)} & \mathsf{N}^{(1)} \\
    \mathsf{N}^{(2)} & \mathsf{D}^{(2)}
  \end{pmatrix}.
\end{equation}

The coupling spin torque, $\mathbf{T}_{\ell}^{(2){\rm SHE}}$, is defined as [see also Eq. (\ref{eq:STT_2})], 
\begin{equation}
  \mathbf{T}_{\ell}^{(2){\rm SHE}}
  =
  \frac{\gamma}{Md_{\rm F}}
  \mathbf{m}_{\ell}
  \times
  \left[
    \mathbf{J}_{\rm s(SHE)}^{{\rm F}_{\ell}/{\rm N}^{\prime}}
    \times
    \mathbf{m}_{\ell}
  \right],
  \label{eq:STT_2_revised}
\end{equation}
where the $k$ ($k=1,2,3$ or $x$,$y$,$z$) component of $\mathbf{J}_{\rm s(SHE)}^{{\rm F}_{\ell}/{\rm N}^{\prime}}$ is given by 
\begin{equation}
\begin{split}
  \mathbf{e}_{k}
  \cdot
  \mathbf{J}_{\rm s(SHE)}^{{\rm F}_{1}/{\rm N}^{\prime}}
  =
  \sum_{a=1}^{3}
  &
  \frac{(1-p_{g}^{\prime 2})g^{\prime}}{4\pi S^{\prime}}
  e \vartheta^{*}
  \lambda_{\rm F}
  E_{x}
\\
  &
  \times
  \left(
    M_{k,a}^{-1}
    m_{1y}
    m_{1a}
    +
    M_{k,a+3}^{-1}
    m_{2y}
    m_{2a}
  \right),
\end{split}
\end{equation}
\begin{equation}
\begin{split}
  \mathbf{e}_{k}
  \cdot
  \mathbf{J}_{\rm s(SHE)}^{{\rm F}_{2}/{\rm N}^{\prime}}
  =
  \sum_{a=1}^{3}
  &
  \frac{(1-p_{g}^{\prime 2})g^{\prime}}{4\pi S^{\prime}}
  e \vartheta^{*}
  \lambda_{\rm F}
  E_{x}
\\
  &\times
  \left(
    M_{k+3,a}^{-1}
    m_{1y}
    m_{1a}
    +
    M_{k+3,a+3}^{-1}
    m_{2y}
    m_{2a}
  \right).
\end{split}
\end{equation}

On the other hand, the matrix $\mathsf{L}$ consists of two contributions as 
$\mathsf{L}=\mathsf{L}_{0}+\mathsf{L}^{\prime}$, 
where $\mathsf{L}_{0}$ is given by 
\begin{equation}
\begin{split}
  \mathsf{L}_{0}
  &=
  \hat{I}
\\
  &+
  \left(
    \alpha
    +
    \alpha^{\prime}
  \right)
  \begin{pmatrix}
    0 & m_{1z} & -m_{1y} & 0 & 0 & 0 \\
    -m_{1z} & 0 & m_{1x} & 0 & 0 & 0 \\
    m_{1y} & -m_{1x} & 0 & 0 & 0 & 0 \\
    0 & 0 & 0 & 0 & m_{2z} & -m_{2y} \\
    0 & 0 & 0 & -m_{2z} & 0 & m_{2x} \\
    0 & 0 & 0 & m_{2y} & -m_{2x} & 0
  \end{pmatrix},
  \label{eq:matrix_L_zero}
\end{split}
\end{equation}
where $\hat{I}$ is the $6\times 6$ unit matrix. 
The parameter $\alpha^{\prime}$ is defined by Eq. (\ref{eq:alpha_prime}). 
We emphasize that the spin pumping from the ferromagnet emits pure spin current 
not only to the nonmagnetic connector N${}^{\prime}$ but also to the bottom nonmagnet N, 
although the discussion in Sec. \ref{sec:Spin pumping} mainly focuses on the former effect only. 
The spin pumping to the bottom nonmagnet results in an enhancement of the damping constant [\onlinecite{tserkovnyak02a}]. 
Since the thickness of the bottom nonmagnet, $d_{\rm N}=3$ nm, is larger than the spin diffusion length $\lambda_{\rm N}=1.2$ nm, 
we neglect the backflow [\onlinecite{tserkovnyak02b}] from the bottom nonmagnet, for simplicity. 
Then, the enhancement of the damping constant due to the spin pumping into the bottom nonmagnet is given by $\alpha^{\prime}$. 
This is the origin of $\alpha^{\prime}$ in Eq. (\ref{eq:matrix_L_zero}). 

On the other hand, the components of a $6 \times 6$ matrix $\mathsf{L}^{\prime}$ are given by 
\begin{equation}
\begin{split}
  L_{i,j}^{\prime}
  =
  -\alpha^{\prime}
  &
  \left[
    \left(
      M_{i,a}^{-1}
      m_{1b}
      -
      M_{i,b}^{-1}
      m_{1a}
    \right)
  \right.
\\
  &-
    m_{1x}
    m_{1i}
    \left(
      M_{1,a}^{-1}
      m_{1b}
      -
      M_{1,b}^{-1}
      m_{1a}
    \right)
\\
  &-
    m_{1y}
    m_{1i}
    \left(
      M_{2,a}^{-1}
      m_{1b}
      -
      M_{2,b}^{-1}
      m_{1a}
    \right)
\\
  &-
  \left.
    m_{1z}
    m_{1i}
    \left(
      M_{3,a}^{-1}
      m_{1b}
      -
      M_{3,b}^{-1}
      m_{1a}
    \right)
  \right],
\end{split}
\end{equation}
\begin{equation}
\begin{split}
  L_{i,j+3}^{\prime}
  =
  -\alpha^{\prime}
  &
  \left[
    \left(
      M_{i,a+3}^{-1}
      m_{2b}
      -
      M_{i,b+3}^{-1}
      m_{2a}
    \right)
  \right.
\\
  &-
    m_{1x}
    m_{1i}
    \left(
      M_{1,a+3}^{-1}
      m_{2b}
      -
      M_{1,b+3}^{-1}
      m_{2a}
    \right)
\\
  &-
    m_{1y}
    m_{1i}
    \left(
      M_{2,a+3}^{-1}
      m_{2b}
      -
      M_{2,b+3}^{-1}
      m_{2a}
    \right)
\\
  &-
  \left.
    m_{1z}
    m_{1i}
    \left(
      M_{3,a+3}^{-1}
      m_{2b}
      -
      M_{3,b+3}^{-1}
      m_{2a}
    \right)
  \right],
\end{split}
\end{equation}
\begin{equation}
\begin{split}
  L_{i+3,j}^{\prime}
  =
  -\alpha^{\prime}
  &
  \left[
    \left(
      M_{i+3,a}^{-1}
      m_{1b}
      -
      M_{i+3,b}^{-1}
      m_{1a}
    \right)
  \right.
\\
  &-
    m_{2x}
    m_{2i}
    \left(
      M_{4,a}^{-1}
      m_{1b}
      -
      M_{4,b}^{-1}
      m_{1a}
    \right)
\\
  &-
    m_{2y}
    m_{2i}
    \left(
      M_{5,a}^{-1}
      m_{1b}
      -
      M_{5,b}^{-1}
      m_{1a}
    \right)
\\
  &-
  \left.
    m_{2z}
    m_{2i}
    \left(
      M_{6,a}^{-1}
      m_{1b}
      -
      M_{6,b}^{-1}
      m_{1a}
    \right)
  \right],
\end{split}
\end{equation}
\begin{equation}
\begin{split}
  L_{i+3,j+3}^{\prime}
  =
  -\alpha^{\prime}
  &
  \left[
    \left(
      M_{i+3,a+3}^{-1}
      m_{2b}
      -
      M_{i+3,b+3}^{-1}
      m_{2a}
    \right)
  \right.
\\
  &-
    m_{2x}
    m_{2i}
    \left(
      M_{4,a+3}^{-1}
      m_{2b}
      -
      M_{4,b+3}^{-1}
      m_{2a}
    \right)
\\
  &-
    m_{2y}
    m_{2i}
    \left(
      M_{5,a+3}^{-1}
      m_{2b}
      -
      M_{5,b+3}^{-1}
      m_{2a}
    \right)
\\
  &-
  \left.
    m_{2z}
    m_{2i}
    \left(
      M_{6,a+3}^{-1}
      m_{2b}
      -
      M_{6,b+3}^{-1}
      m_{2a}
    \right)
  \right],
\end{split}
\end{equation}
where $i,j=1,2,3$ or $x$,$y$,$z$, 
whereas $(a,b)=(2,3)$ for $j=1$, $(3,1)$ for $j=2$, and $(1,2)$ for $j=3$. 
The parameter $\alpha^{\prime}$ is given by 
\begin{equation}
  \alpha^{\prime}
  =
  \frac{\gamma \hbar g_{\rm r}^{\prime}}{4\pi MS^{\prime}d_{\rm F}}. 
  \label{eq:alpha_prime}
\end{equation}
For F${}_{\ell}$/N${}^{\prime}$ interface, we use the values of the parameters used in Ref. [\onlinecite{taniguchi18PRB}], 
i.e., $p_{g}=0.50$, $r^{\prime}=0.25 $k$\Omega$nm${}^{2}$, and $g_{\rm r}^{\prime}/S^{\prime}=15$ nm${}^{-2}$. 
As a result, $\alpha^{\prime}$ becomes $0.0074$. 
On the other hand, $\alpha^{\prime\prime}$ in Eq. (\ref{eq:alpha_pp}) is $0.0031$.
For the bulk parameter of the nonmagnetic connector N${}^{\prime}$, 
we use $1/\sigma_{\rm N^{\prime}}=21$ $\Omega$nm and $\lambda_{\rm N^{\prime}}=500$ nm, which is a typical value of the spin diffusion length in Cu [\onlinecite{bass07}]. 
The off-diagonal components of $\mathsf{L}^{\prime}$, 
$L_{i,j+3}^{\prime}$ and $L_{i+3,j}^{\prime}$, lead to the coupled motion of the magnetizations. 

% ===================================================================================================================================================================================== %

% ===================================================================================================================================================================================== %

\begin{figure}%[p]
\centerline{\includegraphics[width=1.0\columnwidth]{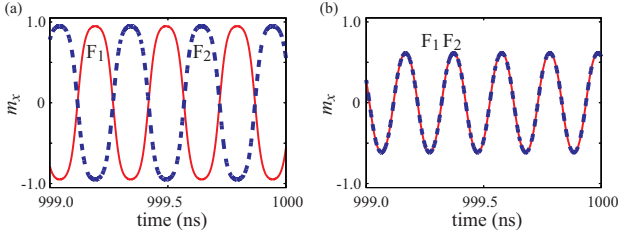}}%\vspace{-3.0ex}
\caption{
         (a) Time evolutions of $m_{1x}$ and $m_{2x}$ in the presence of the coupling torque given by Eq. (\ref{eq:STT_2_revised}).
             The current density is $J_{0}=28$ MA/cm${}^{2}$. 
             The spin pumping is neglected. 
         (b) The time evolutions of $m_{1x}$ and $m_{2x}$ in the presence of the spin pumping.
             The current density is $J_{0}=45$ MA/cm${}^{2}$. 
             The coupling spin torque given by Eq. (\ref{eq:STT_2_revised}) is neglected. 
         \vspace{-3ex}}
\label{fig:fig8}
\end{figure}

% ===================================================================================================================================================================================== %

% ===================================================================================================================================================================================== %

We note that Eq. (\ref{eq:STT_2_revised}) is a revised formula of Eq. (\ref{eq:STT_2}), where the interface effect is newly included. 
Figure \ref{fig:fig8}(a) shows an example of the magnetization dynamics in the presence of the coupling torque given by Eq. (\ref{eq:STT_2_revised}), 
where the spin pumping is neglected by setting $\alpha^{\prime}=\alpha^{\prime\prime}=0$ and the current density is $J_{0}=28$ MA/cm${}^{2}$. 
It should be emphasized that the antiphase synchronization is excited between $\mathbf{m}_{1}$ and $\mathbf{m}_{2}$, 
which is consistent with the results obtained by using Eq. (\ref{eq:STT_2}) [see Fig. \ref{fig:fig4}(a)]. 
We note that the oscillation frequency of $m_{\ell x}$ shown in Fig. \ref{fig:fig8}(a) is slightly different from that shown in Fig. \ref{fig:fig4}(a), 
where the current density is identical. 
The difference is considered as due to the fact that the strengths of the coupling torque given by Eqs. (\ref{eq:STT_2}) and (\ref{eq:STT_2_revised}) are slightly different 
because of the presence of the interface parameter and the spin diffusion length of the connector in Eq. (\ref{eq:STT_2_revised}). 

% ===================================================================================================================================================================================== %

On the other hand, Fig. \ref{fig:fig8}(b) shows the coupled dynamics of the magnetizations via spin pumping, 
where the coupling spin torque given by Eq. (\ref{eq:STT_2_revised}) is neglected. 
The current magnitude, $J_{0}=45$ MA/cm${}^{2}$, is large compared with that used in Fig. \ref{fig:fig8}(a). 
This is because the enhancement of the damping constant due to the spin pumping leads to an increase of the critical current density, as mentioned in the main text. 
The result shown in Fig. \ref{fig:fig8}(b) indicates that the spin pumping prefers the in-phase synchronization between STOs, 
which is consistent with the results shown in Ref. [\onlinecite{taniguchi18PRB}].

% ===================================================================================================================================================================================== %

\subsection{Validity of Eq. (\ref{eq:STT_2})}

One may be interested in addressing whether Eq. (\ref{eq:STT_2}) is a well-approximated formula of Eq. (\ref{eq:STT_2_revised}). 
To answer this question, let us show another approach to calculate the spin torque $\mathbf{T}_{\ell}^{(2){\rm SHE}}$ given by Eq. (\ref{eq:STT_2}). 
Substituting Eq. (\ref{eq:spin_accumulation_N_p_appendix_spin_pumping}) into Eq. (\ref{eq:spin_current_FN_appendix_2}), 
we notice that the spin current at the F${}_{1}$/N${}^{\prime}$ interface is determined by the following equation;
\begin{equation}
\begin{split}
  &
  \frac{\mathbf{J}_{\rm s}^{{\rm F}_{1}/{\rm N}^{\prime}}}{\eta \mathcal{S}}
  -
  \frac{\xi(1-\eta \mathcal{S})}{\eta \mathcal{S}}
  \left(
    \mathbf{m}_{1}
    \cdot
    \mathbf{J}_{\rm s}^{{\rm F}_{1}/{\rm N}^{\prime}}
  \right)
  \mathbf{m}_{1}
  +
  \frac{\mathbf{J}_{\rm s}^{{\rm F}_{2}/{\rm N}^{\prime}}}{\mathcal{S}}
\\
  &
  -
  \frac{\xi}{\mathcal{S}}
  \left(
    \mathbf{m}_{1}
    \cdot
    \mathbf{J}_{\rm s}^{{\rm F}_{2}/{\rm N}^{\prime}}
  \right)
  \mathbf{m}_{1}
  =
  \mathbf{J}_{\rm s}^{(1)}, 
  \label{eq:current_equation_F1}
\end{split}
\end{equation}
where we introduce $\eta=g_{\rm r}^{\prime}/[g_{\rm N^{\prime}}\sinh(L/\lambda_{\rm N^{\prime}})+g_{\rm r}^{\prime}\cosh(L/\lambda_{\rm N^{\prime}})]$, 
$\mathcal{S}=g_{\rm N^{\prime}}\sinh(L/\lambda_{\rm N^{\prime}})/g_{\rm r}^{\prime}$, and $\xi=1-[(1-p_{g}^{\prime\ 2})g^{\prime}]/(2g_{\rm r}^{\prime})]$, according to Ref. [\onlinecite{chiba15}]. 
We obtain another equation similar to Eq. (\ref{eq:current_equation_F1}) by focusing on the F${}_{2}$/N${}^{\prime}$ interface 
and reversing the suffixes $1$ and $2$. 
Solving these equations in parallel with respect to $\mathbf{J}_{\rm s}^{{\rm F}_{1}/{\rm N}^{\prime}}$ and $\mathbf{J}_{\rm s}^{{\rm F}_{2}/{\rm N}^{\prime}}$, 
we find that 
\begin{equation}
\begin{split}
  \mathbf{J}_{\rm s}^{{\rm F}_{\ell}/{\rm N}^{\prime}}
  =&
  \frac{\eta \mathcal{S}}{1-\eta^{2}}
  \mathbf{J}_{\rm s}^{(\ell)}
  -
  \frac{\eta^{2} \mathcal{S}}{1-\eta^{2}}
  \mathbf{m}_{\ell}
  \times
  \left[
    \mathbf{J}_{\rm s}^{(\ell^{\prime})}
    \times
    \mathbf{m}_{\ell}
  \right]
\\
  &+
  C_{\ell}
  \mathbf{m}_{\ell}
  \times
  \left(
    \mathbf{m}_{\ell^{\prime}}
    \times
    \mathbf{m}_{\ell}
  \right)
  +
  D_{\ell}
  \mathbf{m}_{\ell},
  \label{eq:current_solution}
\end{split}
\end{equation}
where $(\ell,\ell^{\prime})=(1,2)$ or $(2,1)$. 
The coefficients $C_{\ell}$ and $D_{\ell}$ are determined by the following equations; 
\begin{equation}
\begin{split}
  &
  -\frac{\tilde{z}}{\eta}
  C_{\ell}
  +
  \frac{1-\xi(1-\eta \mathcal{S})}{\eta}
  D_{\ell}
  +
  \left[
    1
    -
    \xi
    \left(
      1
      -
      \tilde{z}^{2}
    \right)
  \right]
  C_{\ell^{\prime}}
  -
  \xi
  \tilde{z}
  D_{\ell^{\prime}}
\\
  &=
  \frac{\eta \mathcal{S}}{1-\eta^{2}}
  \left[
    \frac{\xi(1-\eta^{2}-\eta \mathcal{S})}{\eta}
    \mathbf{m}_{\ell}
    \cdot
    \mathbf{J}_{\rm s}^{(\ell)}
    -
    \left(
      1
      -
      \xi
    \right)
    \mathbf{m}_{\ell}
    \cdot
    \mathbf{J}_{\rm s}^{(\ell^{\prime})}
  \right.
\\
  &
  \left.
    +
    \xi
    \eta
    \tilde{z}
    \mathbf{m}_{\ell^{\prime}}
    \cdot
    \mathbf{J}_{\rm s}^{(\ell)}
  \right],
\end{split}
\end{equation}
\begin{equation}
  \frac{C_{\ell}}{\eta}
  -
  \tilde{z}
  C_{\ell^{\prime}}
  +
  D_{\ell^{\prime}}
  =
  -\frac{\eta^{2} \mathcal{S}}{1-\eta^{2}}
  \mathbf{m}_{\ell^{\prime}}
  \cdot
  \mathbf{J}_{\rm s}^{(\ell^{\prime})},
\end{equation}
where $\tilde{z}=\mathbf{m}_{1}\cdot\mathbf{m}_{2}$. 
Although the general solutions of $C_{\ell}$ and $D_{\ell}$ are complex, 
we note that Eq. (\ref{eq:current_solution}) reproduces the results in Ref. [\onlinecite{chiba15}] 
in the limit of $\xi \to 1$ and $\mathbf{m}_{\ell}\cdot\mathbf{J}_{\rm s}^{(\ell)}=0$, 
where $C_{\ell}=-\eta^{3} \mathcal{S} \{[\mathbf{m}_{\ell^{\prime}}\cdot\mathbf{J}_{\rm s}^{(\ell)}]+ \eta \tilde{z} [\mathbf{m}_{\ell}\cdot\mathbf{J}_{\rm s}^{(\ell^{\prime})}]\}/[(1-\eta^{2})(1-\eta^{2}\tilde{z}^{2})]$ and $D_{\ell}=0$. 
The results imply that the terms related to $C_{\ell}$ and $D_{\ell}$ are 
higher order terms of a small parameter $\eta$ ($0<\eta<1$). 

Now let us consider the coupling spin torque originated from the spin accumulation. 
In this respect, we neglect the spin pumping effect from $\mathbf{J}_{\rm s}^{(\ell)}$. 
Then, the first term on the right hand side of Eq. (\ref{eq:current_solution}) does not contribute to the coupling torque 
because $\mathbf{J}_{\rm s}^{(\ell)}$ is parallel to $\mathbf{m}_{\ell}$. 
On the other hand, using Eq. (\ref{eq:spin_current_SHE}), the coupling spin torque $[\gamma/(Md_{\rm F})]\mathbf{m}_{\ell}\times(\mathbf{J}_{\rm s}^{{\rm F}_{\ell}/{\rm N}^{\prime}}\times\mathbf{m}_{\ell})$ 
which contributes from the second term of Eq. (\ref{eq:current_solution}) becomes 
\begin{equation}
  -\frac{\gamma \hbar \tilde{\vartheta}^{\prime} J_{0}}{2eMd_{\rm F}} 
  m_{\ell^{\prime}y}
  \mathbf{m}_{\ell}
  \times
  \left(
    \mathbf{m}_{\ell^{\prime}}
    \times
    \mathbf{m}_{\ell}
  \right),
  \label{eq:STT_2_exact_approx}
\end{equation}
where $\tilde{\vartheta}^{\prime}$ is 
\begin{equation}
  \tilde{\vartheta}^{\prime}
  =
  \frac{\eta^{2}S}{1-\eta^{2}}
  \frac{(1-p_{g}^{\prime\ 2})g^{\prime}e^{2} \lambda_{\rm F}}{2\pi S \hbar \sigma_{\rm N}}
  \vartheta^{*}. 
  \label{eq:coupling_constant_revised}
\end{equation}
Equation (\ref{eq:STT_2_exact_approx}) indicates that the exact solution of the coupling spin torque, 
$\mathbf{T}_{\ell}^{(2){\rm SHE}}$, given by Eq. (\ref{eq:STT_2_revised}), definitely provides a torque 
having the same angular dependence and sign with Eq. (\ref{eq:STT_2}). 
The parameter $\tilde{\vartheta}^{\prime}$ given by Eq. (\ref{eq:coupling_constant_revised}) corresponds to Eq. (\ref{eq:coupling_constant}), 
where the interface effect is included in Eq. (\ref{eq:coupling_constant_revised}). 

% ================================================================================================================================================================================= %

\section{Conditions of numerical simulations}
\label{sec:AppendixB}

We use the same method developed in Ref. [\onlinecite{taniguchi17}] to evaluate the phase difference of the synchronized oscillators, 
where the phase difference is determined from the oscillation period, as in the case of the Kuramoto model. 
Therefore, the in-phase and antiphase correspond to 0 and 0.5 of the vertical axis in Fig. \ref{fig:fig7}. 
We solve the LLG equation from $t=0$ to $t=1.0$ $\mu$s with $N_{\rm t}=10^{8}$ time mesh, 
and gather $N_{\rm i}=2^{26}=67108864$ data of $\mathbf{m}_{\ell}$ ($\ell=1,2$) from $t=(N_{\rm t}-N_{\rm i}+1)\Delta t$ to $t=N_{\rm t}\Delta t=1$ $\mu$s, 
where $\Delta t=1.0\mu {\rm s}/N_{\rm t}=10$ fs. 
Therefore, the frequency step of the Fourier transformation becomes $1/(N_{\rm i}\Delta t)=1.5$ MHz. 

% ================================================================================================================================================================================= %

% ================================================================================================================================================================================= %

\section{Oscillation trajectory of in-plane magnetized ferromagnet}
\label{sec:AppendixC}

In principle, an analytical solution of the oscillation trajectory of the magnetization can be obtained by solving Eq. (\ref{eq:LLG}). 
However, the LLG equation is a nonlinear equation, and therefore, it is usually difficult to obtain an analytical solution. 
If we focus on an self-oscillation state an approximate solution of the oscillation trajectory can nevertheless be obtained. 
The self-oscillation state is excited when the dissipation due to the damping torque balances the work done by the spin torque. 
In this case, the magnetization can be approximated as moving on a constant energy curve of $E=-M \int d \mathbf{m}\cdot\mathbf{H}$. 
The oscillation trajectory on the constant energy curve is obtained from the Landau-Lifshitz equation $d \mathbf{m}/dt=-\gamma\mathbf{m}\times\mathbf{H}$, 
and is given by 
\begin{equation}
  m_{x}
  =
  \sqrt{
    1
    +
    \frac{2E}{MH_{\rm K}}
  }
  {\rm sn}
  \left[
    \frac{4 \mathsf{K}(k)}{\tau(E)}
    t
    +
    \varphi_{0},
    k
  \right],
  \label{eq:mx}
\end{equation}
\begin{equation}
  m_{y}
  =
  \sqrt{
    \frac{4\pi M-2E/M}{H_{\rm K}+4\pi M}
  }
  {\rm dn}
  \left[
    \frac{4 \mathsf{K}(k)}{\tau(E)}
    t
    +
    \varphi_{0},
    k
  \right],
  \label{eq:my}
\end{equation}
\begin{equation}
  m_{z}
  =
  \sqrt{
    \frac{H_{\rm K}+2E/M}{H_{\rm K}+4\pi M}
  }
  {\rm cn}
  \left[
    \frac{4 \mathsf{K}(k)}{\tau(E)}
    t
    +
    \varphi_{0},
    k
  \right],
  \label{eq:mz}
\end{equation}
where ${\rm sn}(u,k)$, ${\rm dn}(u,k)$, and ${\rm cn}(u,k)$ are the Jacobi elliptic functions, 
whereas $\mathsf{K}(k)$ is the first kind of complete elliptic integral. 
The modulus of the elliptic function and integral is 
\begin{equation}
  k
  =
  \sqrt{
    \frac{4\pi M(H_{\rm K}+2E/M)}{H_{\rm K}(4\pi M-2E/M)}
  }.
  \label{eq:moduls}
\end{equation}
The oscillation period $\tau(E)$ is related to the frequency of the self-oscillation $f(E)$ via $f(E)=1/\tau(E)$, where 
\begin{equation}
  f(E)
  =
  \frac{\gamma \sqrt{H_{\rm K}(4\pi M-2E/M)}}{4 \mathsf{K}(k)}.
\end{equation}
The Jacobi elliptic functions can be expanded as an infinite Fourier series [\onlinecite{byrd71}]. 
Therefore, the oscillation trajectory in the real space cannot be described by trigonometric functions with a single frequency, in general. 
In the small amplitude limit, however, $m_{x}$ and $m_{z}$ are well described by $\sin(2\pi f_{\rm FMR}t+\varphi_{0})$ and $\cos(2\pi f_{\rm FMR}t+\varphi_{0})$ 
with the FMR frequency $f_{\rm FMR}=\gamma\sqrt{H_{\rm K}(H_{\rm K}+4\pi M)}/(2\pi)$, whereas $m_{y}$ becomes almost constant. 
This conclusion can be confirmed by notifying the fact that the small amplitude limit corresponds to $k \to 0$, 
and therefore, ${\rm sn}(u,k)\to \sin u$, ${\rm cn}(u,k) \to \cos u$, and ${\rm dn}(u,k) \to 1$.

% ================================================================================================================================================================================= %

% ===================================================================================================================================================================================== %

\section{Scaling currents}
\label{sec:AppendixD}

In this Appendix, we show how the coupling torque $\tilde{\mathbf{T}}_{\ell}^{(2){\rm SHE}}$ between two oscillators affects the critical and switching current densities. 

% ===================================================================================================================================================================================== %

\subsection{Definition of critical and switching currents}

First, let us briefly review the derivation of Eqs. (\ref{eq:J_c}) and (\ref{eq:J_star}). 
The self-oscillation is excited when the dissipation due to the damping torque balances the work done by the spin torque. 
This condition is expressed as 
\begin{equation}
  \oint 
  dt 
  \frac{dE}{dt}
  =
  0,
  \label{eq:dEdt}
\end{equation}
where $dE/dt=-M \mathbf{H}_{\ell}\cdot(d \mathbf{m}_{\ell}/dt)$. 
Note that $d \mathbf{m}_{\ell}/dt$ is given by the LLG equation. 
The integral of Eq. (\ref{eq:dEdt}) should be performed on a constant energy curve. 
Let us denote the current satisfying Eq. (\ref{eq:dEdt}) as $J(E)$, 
which is a function of the energy density $E$. 
The critical and switching current densities given by Eqs. (\ref{eq:J_c}) and (\ref{eq:J_star}) can be defined as [\onlinecite{hillebrands06}] 
\begin{equation}
  J_{\rm c}
  =
  \lim_{E \to E_{\rm min}}
  J(E),
\end{equation}
\begin{equation}
  J^{*}
  =
  \lim_{E \to E_{\rm saddle}}
  J(E),
\end{equation}
where $E_{\rm min}=-MH_{\rm K}/2$ and $E_{\rm saddle}=0$ correspond to the minimum and saddle point energies, respectively. 
The purpose of this Appendix is to derive the explicit forms of $J_{\rm c}$ and $J^{*}$ in the presence of the coupling. 

% ===================================================================================================================================================================================== %

\subsection{A general guideline to calculate $J(E)$}

According to Eq. (\ref{eq:dEdt}), the calculation of $J(E)$ requires to perform the following types of integral, 
\begin{equation}
  \oint 
  dt 
  F(\mathbf{m}_{\ell},\mathbf{m}_{\ell^{\prime}}), 
  \label{eq:integral_def}
\end{equation}
where $F$ is an arbitrary function of $\mathbf{m}_{\ell}$ and $\mathbf{m}_{\ell^{\prime}}$. 
Let us first give a general direction to perform this kind of integral. 

As mentioned above, the magnetization in the self-oscillation state can be approximated as precessing on a constant energy curve. 
The trajectory of the constant energy curve is described by Eqs. (\ref{eq:mx})-(\ref{eq:mz}) in Appendix \ref{sec:AppendixC}, 
where the initial state is determined by the phase $\varphi_{0}$. 
For convention, in this section, we set $\varphi_{0}$ of $\mathbf{m}_{\ell}$ to be zero, 
whereas that of $\mathbf{m}_{\ell^{\prime}}$ as $\Delta\varphi$. 
Here, $\Delta\varphi$ can be regarded as the phase difference between $\mathbf{m}_{\ell}$ and $\mathbf{m}_{\ell^{\prime}}$. 
In principle, the integral, Eq. (\ref{eq:integral_def}), can be calculated by substituting the solutions of $\mathbf{m}_{\ell}$ and $\mathbf{m}_{\ell^{\prime}}$ into Eq. (\ref{eq:integral_def}) 
and using the integral formulas of the elliptic functions [\onlinecite{byrd71}]. 
There is, however, another approach to calculate the integral. 
Let us introduce new variable $x$ as $x={\rm sn}(u,k)$, where $u=4 \mathsf{K}(k) t/\tau(E)$. 
Then, we find that $du=dx/\sqrt{(1-x^{2})(1-k^{2}x^{2})}$. 
Therefore, the integral becomes 
\begin{equation}
  \oint 
  dt 
  =
  \frac{4}{\gamma \sqrt{H_{\rm K}(4\pi M-2E/M)}}
  \int_{0}^{1}
  \frac{dx}{\sqrt{(1-x^{2})(1-k^{2}x^{2})}}. 
\end{equation}
The numerical factor $4$ appears because of the symmetry, 
i.e., we perform the integral over the time range of $0 \le t \le \tau/4$, and multiply the numerical factor $4$. 
This simplification is allowed due to the fact that the work done by the spin torque and the dissipation due to the damping torque 
during the time $0 \le t \le \tau/4$ is independent of the choice of the initial conditions. 
We note that the other elliptic functions can be expressed in terms of $x$ as ${\rm cn}(u,k)=\sqrt{1-x^{2}}$ and ${\rm dn}(u,k)=\sqrt{1-k^{2}x^{2}}$. 
Therefore, the integral, Eq. (\ref{eq:integral_def}), can be rewritten as 
\begin{equation}
\begin{split}
  &
  \int
  \frac{dx}{\sqrt{(1-x^{2})(1-k^{2}x^{2})}}
  F(x)
\\
  &=
  \int
  dx 
%  \sum_{a=-\infty}^{\infty}
%  \sum_{b=-\infty}^{\infty}
%  \sum_{c=-\infty}^{\infty}
  x^{a} 
  \left(
    1
    -
    x^{2}
  \right)^{b/2}
  \left(
    1
    -
    k^{2}
    x^{2}
  \right)^{c/2}, 
  \label{eq:integral_general}
\end{split}
\end{equation}
where $a$, $b$, and $c$ are some numbers determined by the explicit form of $F(x)$. 
The examples of the integral will appear below. 

% ===================================================================================================================================================================================== %

\subsection{Critical and switching current densities of coupled two oscillators}

In the present system, the left hand side of Eq. (\ref{eq:dEdt}) consists of three contributions; 
the works done by the spin torques, $\mathbf{T}_{\ell}^{(1)}$ and $\tilde{\mathbf{T}}_{\ell}^{(2){\rm SHE}}$, and 
the dissipation due to the damping torque. 
The work done by the spin torque $\mathbf{T}_{\ell}^{(1)}$ and 
the dissipation due to the damping torque in the present case are, respectively, given by 
\begin{equation}
\begin{split}
  \mathscr{W}_{\rm s1}
  &=
  \oint
  dt 
  \gamma 
  M 
  H_{\rm s1}
  \left[
    \mathbf{e}_{y}
    \cdot
    \mathbf{H}_{\ell}
    -
    \left(
      \mathbf{m}_{\ell}
      \cdot
      \mathbf{e}_{y}
    \right)
    \left(
      \mathbf{m}_{\ell}
      \cdot
      \mathbf{H}_{\ell}
    \right)
  \right]
\\
  &=
  \frac{\pi \hbar \vartheta_{\rm R} J_{0} (H_{\rm K}+2E/M)}{ed_{\rm F} \sqrt{H_{\rm K} (H_{\rm K} + 4\pi M)}}.
  \label{eq:W_s1}
\end{split}
\end{equation}
\begin{equation}
\begin{split}
  \mathscr{W}_{\alpha}
  &=
  -\oint 
  dt 
  \alpha 
  \gamma 
  M 
  \left[
    \mathbf{H}_{\ell}^{2}
    -
    \left(
      \mathbf{m}_{\ell}
      \cdot
      \mathbf{H}_{\ell}
    \right)^{2}
  \right]
\\
  &=
  -4 \alpha M 
  \sqrt{
    \frac{4\pi M-2E/M}{H_{\rm K}}
  }
  \left[
    \frac{2E}{M}
    \mathsf{K}(k)
    +
    H_{\rm K}
    \mathsf{E}(k)
  \right],
  \label{eq:W_alpha}
\end{split}
\end{equation}
where $\mathsf{E}(k)$ is the second kind of complete elliptic integral, 
whereas $H_{\rm s1}=\hbar \vartheta_{\rm R}J_{0}/(2eMd_{\rm F})$.  

The definition of the work done by the coupling torque $\tilde{\mathbf{T}}_{\ell}^{(2){\rm SHE}}$ is defined as 
\begin{equation}
  \mathscr{W}_{\rm s2}
  =
  \oint 
  dt 
  \gamma
  M 
  H_{\rm s2}
  m_{\ell^{\prime}y}
  \left[
    \mathbf{m}_{\ell^{\prime}}
    \cdot
    \mathbf{H}_{\ell}
    -
    \left(
      \mathbf{m}_{\ell}
      \cdot
      \mathbf{m}_{\ell^{\prime}}
    \right)
    \left(
      \mathbf{m}_{\ell}
      \cdot
      \mathbf{H}_{\ell}
    \right)
  \right].
  \label{eq:W_s2_def}
\end{equation}
It is difficult to calculate $\mathscr{W}_{\rm s2}$ for an arbitrary phase difference $\Delta\varphi$. 
The numerical simulation indicates that the antiphase synchronization is stable for the coupled two STOs. 
Therefore, we focus on the case of the antiphase, $\Delta\varphi=2 \mathsf{K}(k)$. 
Note that the elliptic functions satisfy 
${\rm sn}[u+2\mathsf{K}(k),k]=-{\rm sn}(u,k)$,
${\rm cn}[u+2\mathsf{K}(k),k]=-{\rm cn}(u,k)$, and 
${\rm dn}[u+2\mathsf{K}(k),k]={\rm dn}(u,k)$. 
Then, we find that 
\begin{equation}
\begin{split}
  &
  \mathscr{W}_{\rm s2}(\Delta\varphi=2\mathsf{K}(k))
\\
  &=
  2 \gamma 
  M H_{\rm s2}
  \oint 
  dt 
  \left[
    H_{\rm K}
    \left(
      m_{\ell y}^{3}
      -
      m_{\ell y}^{5}
    \right)
    +
    4\pi M 
    m_{\ell y}^{3}
    m_{\ell z}^{2}
  \right]. 
  \label{eq:W_s2_antiphase}
\end{split}
\end{equation}
The integrals on the right hand side are calculated by using Eqs. (\ref{eq:mx})-(\ref{eq:mz}) as 
\begin{widetext}
\begin{equation}
\begin{split}
  \oint 
  dt 
  m_{\ell y}^{3}
  &=
  \frac{4}{\gamma \sqrt{H_{\rm K}(4\pi M-2E/M)}}
  \int_{0}^{1}
  \frac{dx}{\sqrt{(1-x^{2})(1-k^{2}x^{2})}}
  \left(
    \frac{4\pi M-2E/M}{H_{\rm K}+4\pi M}
  \right)^{3/2}
  {\rm dn}^{3}(u,k)
\\
  &=
  \frac{4}{\gamma \sqrt{H_{\rm K}(4\pi M-2E/M)}}
  \left(
    \frac{4\pi M-2E/M}{H_{\rm K}+4\pi M}
  \right)^{3/2}
  \int_{0}^{1}
  dx 
  \frac{1-k^{2}x^{2}}{\sqrt{1-x^{2}}}
\\
  &=
  \frac{4}{\gamma \sqrt{H_{\rm K}(4\pi M-2E/M)}}
  \left(
    \frac{4\pi M-2E/M}{H_{\rm K}+4\pi M}
  \right)^{3/2}
  \left[
    \frac{k^{2}x \sqrt{1-x^{2}}}{2}
    +
    \frac{2-k^{2}}{2}
    \sin^{-1}x
  \right]
  \bigg|_{0}^{1}
\\
  &=
  \frac{4}{\gamma \sqrt{H_{\rm K}(4\pi M-2E/M)}}
  \left(
    \frac{4\pi M-2E/M}{H_{\rm K}+4\pi M}
  \right)^{3/2}
  \frac{\pi(2-k^{2})}{4},
\end{split}
\end{equation}

\begin{equation}
\begin{split}
  \oint 
  dt 
  m_{\ell y}^{5}
  &=
  \frac{4}{\gamma \sqrt{H_{\rm K}(4\pi M-2E/M)}}
  \int_{0}^{1}
  \frac{dx}{\sqrt{(1-x^{2})(1-k^{2}x^{2})}}
  \left(
    \frac{4\pi M-2E/M}{H_{\rm K}+4\pi M}
  \right)^{5/2}
  {\rm dn}^{5}(u,k)
\\
  &=
  \frac{4}{\gamma \sqrt{H_{\rm K}(4\pi M-2E/M)}}
  \left(
    \frac{4\pi M-2E/M}{H_{\rm K}+4\pi M}
  \right)^{5/2}
  \int_{0}^{1}
  dx 
  \frac{(1-k^{2}x^{2})^{2}}{\sqrt{1-x^{2}}}
\\
  &=
  \frac{4}{\gamma \sqrt{H_{\rm K}(4\pi M-2E/M)}}
  \left(
    \frac{4\pi M-2E/M}{H_{\rm K}+4\pi M}
  \right)^{5/2}
\\
  &\ \ \ \ \times
  \left\{
    \frac{k^{2}x \sqrt{1-x^{2}} [8-3k^{2}(3+2x^{2})] + (8-8k^{2}+3k^{4}) \sin^{-1}x}{8}
  \right\}
  \bigg|_{0}^{1}
\\
  &=
  \frac{4}{\gamma \sqrt{H_{\rm K}(4\pi M-2E/M)}}
  \left(
    \frac{4\pi M-2E/M}{H_{\rm K}+4\pi M}
  \right)^{5/2}
  \frac{\pi(8-8k^{2}+3k^{4})}{16}.
\end{split}
\end{equation}

\begin{equation}
\begin{split}
  \oint
  dt
  m_{\ell y}^{3}
  m_{\ell z}^{2}
  &=
  \frac{4}{\gamma \sqrt{H_{\rm K}(4\pi M-2E/M)}}
  \int_{0}^{1}
  \frac{dx}{\sqrt{(1-x^{2})(1-k^{2}x^{2})}}
  \left(
    \frac{4\pi M-2E/M}{H_{\rm K}+4\pi M}
  \right)^{3/2}
\\
  &\ \ \ \ \ \ \ \ \ \ \ \ \ \ \ \ \ \ \ \ \ \ \ \ \ \ \ \ \ \ \ \ \ \ \ \ \ \ \times
  \left(
    \frac{H_{\rm K}+2E/M}{H_{\rm K}+4\pi M}
  \right)
  {\rm dn}^{3}(u,k)
  {\rm cn}^{2}(u,k)
\\
  &=
  \frac{4}{\gamma \sqrt{H_{\rm K}(4\pi M-2E/M)}}
  \left(
    \frac{4\pi M-2E/M}{H_{\rm K}+4\pi M}
  \right)^{3/2}
  \left(
    \frac{H_{\rm K}+2E/M}{H_{\rm K}+4\pi M}
  \right)
\\
  &\ \ \ \ \times
  \int_{0}^{1}
  dx
  \sqrt{1-x^{2}}
  \left(
    1
    -
    k^{2}
    x^{2}
  \right)
\\
  &=
  \frac{4}{\gamma \sqrt{H_{\rm K}(4\pi M-2E/M)}}
  \left(
    \frac{4\pi M-2E/M}{H_{\rm K}+4\pi M}
  \right)^{3/2}
  \left(
    \frac{H_{\rm K}+2E/M}{H_{\rm K}+4\pi M}
  \right)
\\
  &\ \ \ \ \times
  \left\{
    \frac{x\sqrt{1-x^{2}} [4+k^{2}(1-2x^{2})] + (4-k^{2})\sin^{-1}x}{8}
  \right\}
  \bigg|_{0}^{1}
\\
  &=
  \frac{4}{\gamma \sqrt{H_{\rm K}(4\pi M-2E/M)}}
  \left(
    \frac{4\pi M-2E/M}{H_{\rm K}+4\pi M}
  \right)^{3/2}
  \left(
    \frac{H_{\rm K}+2E/M}{H_{\rm K}+4\pi M}
  \right)
  \frac{\pi(4-k^{2})}{16}. 
\end{split}
\end{equation}

\end{widetext}

Summarizing these integrals, we find that 
\begin{equation}
\begin{split}
  \mathscr{W}_{\rm s2}
  =&
  \frac{\pi \hbar \tilde{\vartheta} J_{0}}{ed_{\rm F} \sqrt{H_{\rm K}(H_{\rm K}+4\pi M)}}
\\
  & \times
  \frac{(H_{\rm K}+2E/M)[4\pi M(H_{\rm K}-2E/M) - 2H_{\rm K}(2E/M)]}{H_{\rm K}(H_{\rm K}+4\pi M)}. 
  \label{eq:W_s2}
\end{split}
\end{equation}

Equation (\ref{eq:dEdt}) can be expressed as 
$\mathscr{W}_{\rm s1}(E)+\mathscr{W}_{\rm s2}(E)+\mathscr{W}_{\alpha}(E)=0$. 
The current density $J(E)$ is defined as the current $J_{0}$ satisfying this condition. 
Using Eqs. (\ref{eq:W_s1}), (\ref{eq:W_alpha}), and (\ref{eq:W_s2}), 
%and considering the limits of $E \to E_{\rm min}$ and $E \to E_{\rm saddle}$, 
the critical and switching current densities in the presence of the coupling between two STOs are obtained as Eqs. (\ref{eq:J_c_coupled}) and (\ref{eq:J_star_coupled}), respectively. 

% ===================================================================================================================================================================================== %

\subsection{Critical current density in Fig. \ref{fig:fig2}(a)}

It is clear from Eq. (\ref{eq:W_s2_def}) that the work done by the torque $\tilde{\mathbf{T}}_{\ell}^{(2){\rm SHE}}$ is zero 
when the in-phase synchronization $\mathbf{m}_{1}=\mathbf{m}_{2}$ is excited. 
Therefore, the critical current density in Fig. \ref{fig:fig2}(a) is given by Eq. (\ref{eq:J_c}) 
by replacing $\alpha$ with $\alpha+\alpha^{\prime\prime}$. 
In fact, the value of $J_{\rm c}$ with the damping constant $\alpha+\alpha^{\prime\prime}$ is 42 MA/cm${}^{2}$, 
which is consistent with the numerically calculated the critical current density of 44 MA/cm${}^{2}$.

\end{document}